\documentclass[11pt]{article}
\usepackage{amsfonts}
\usepackage{cite}
\usepackage{bbm}
\usepackage{mathrsfs}
\usepackage{amscd}
\usepackage{color}
\usepackage{amsmath,amsfonts,amssymb,amscd}
\usepackage{indentfirst,graphics,epsfig,psfrag}
\input{epsf}
\usepackage{ifpdf}
\usepackage{enumerate}
\usepackage{appendix}
\usepackage{enumerate}

\usepackage{lineno}

\makeatletter \@addtoreset{figure}{section} \makeatother
\makeatletter
\long\def\@makecaption#1#2{%
   \vskip 10\p@
   \setbox\@tempboxa\hbox{{#1}\ \ #2}%
   \ifdim \wd\@tempboxa >\hsize

       {#1}\ \ #2\par
   \else
       \hbox to\hsize{\hfil\box\@tempboxa\hfil}%
   \fi}
\makeatother

\newtheorem{thm}{Theorem}[section]
\newtheorem{cor}[thm]{Corollary}
\newtheorem{lem}[thm]{Lemma}

\newtheorem{con}[thm]{Conjecture}

\newtheorem{pro}[thm]{Proposition}

\newtheorem{definition}{Definition}[section]

\begin{document}
\title{\textbf{A survey of recent results in (generalized) graph entropies}}
\author{
\small  Xueliang Li, Meiqin Wei\\
\small Center for Combinatorics and LPMC-TJKLC\\
\small Nankai University, Tianjin 300071, P.R. China\\
\small E-mails: lxl@nankai.edu.cn; weimeiqin8912@163.com}
\date{}
\maketitle
\begin{abstract}
The entropy of a graph was first introduced by Rashevsky
\cite{Rashevsky} and Trucco \cite{Trucco} to interpret as the
structural information content of the graph and serve as a
complexity measure. In this paper, we first state a number of
definitions of graph entropy measures and generalized graph
entropies. Then we survey the known results about them from the
following three respects: inequalities and extremal properties on
graph entropies, relationships between graph structures, graph
energies, topological indices and generalized graph entropies,
complexity for calculation of graph entropies. Various applications
of graph entropies together with some open problems and conjectures
are also presented for further research.
\\[2mm]
\\[2mm]

\end{abstract}

\section{Introduction}

\label{ch1:sec1} Graph entropy measures play an important role in a
variety of subjects, including information theory, biology,
chemistry, and sociology. It was first introduced by Rashevsky
\cite{Rashevsky} and Trucco \cite{Trucco}. Mowshowitz
\cite{Mowshowitz1,Mowshowitz2,Mowshowitz3,Mowshowitz4} first defined
and investigated the entropy of graphs, and K\"{o}rner \cite{K}
introduced a different definition of graph entropy closely linked to
problems in information and coding theory. In fact, there may be no
``right" one or ``good" one, since what may be useful in one domain
may not be serviceable in another.

Distinct graph entropies have been used extensively to characterize
the structures of graph-based systems in various fields. In these
applications the entropy of a graph is interpreted as the structural
information content of the graph and serves as a complexity measure.
It is worth mentioning that two different approaches to measure the
complexity of graphs have been developed: deterministic and
probabilistic. The deterministic category encompasses the encoding,
substructure count and generative approaches, while the
probabilistic category includes measures that apply an entropy
function to a probability distribution associated with a graph. The
second category is subdivided into intrinsic and extrinsic
subcategories. Intrinsic measures use structural features of a graph
to partition the graph (usually the set of vertices or edges) and
thereby determine a probability distribution over the components of
the partition. Extrinsic measures impose an arbitrary probability
distribution on graph elements. Both of these categories employ the
probability distribution to compute an entropy value. Shannon's
entropy function is most commonly used, but several different
families of entropy functions are also considered.

Actually, three survey papers \cite{Simonyi1,Simonyi2,DM1} on graph
entropy measures were published already. However, \cite{Simonyi1,
Simonyi2} focused narrowly on the properties of K\"{o}rner's entropy
measures and \cite{DM1} provided an overview of the most well-known
graph entropy measures which contains not so many results and only
concepts were preferred. Here we focus on the development of graph
entropy measures and aim to provide a broad overview of the main
results and applications of the most well-known graph entropy
measures.

Now we start our survey by providing some mathematical
preliminaries. Note that all graphs discussed in this chapter are
assumed to be connected.
\begin{definition}
We use $G=(V,E)$ with $|V|<\infty$ and $E\subseteq {V \choose 2}$ to
denote a {\it finite undirected graph}. If $G=(V,E)$, $|V|<\infty$
and $E\subseteq V\times V$, then $G$ is called a {\it finite
directed graph}. We use $\mathcal {G}_{\cup C}$ to denote the set of
finite undirected connected graphs.
\end{definition}

\begin{definition}
Let $G=(V,E)$ be a graph. The quantity $d(v_i)$ is called the {\it
degree} of a vertex $v_i\in V$ where $d(v_i)$ equals the number of
edges $e\in E$ incident with $v_i$. In the following, we simply
denote $d(v_i)$ by $d_i$. If a graph $G$ has $a_i$ vertices of
degree $\mu_i\ (i=1,2,\cdots,t)$, where
$\Delta(G)=\mu_1>\mu_2>\cdots>\mu_t=\delta(G)$ and
$\sum_{i=1}^{t}a_i=n$, we define the {\it degree sequence} of $G$ as
$D(G)=[\mu_1^{a_1},\mu_2^{a_2},\cdots,\mu_t^{a_t}]$. If $a_i=1$, we
use $\mu_i$ instead of $\mu_i^1$ for convenience.
\end{definition}

\begin{definition}
The {\it distance} between two vertices $u,v\in V$, denoted by
$d(u,v)$, is the length of a shortest path between $u,v\in V$. A
path $P$ connecting $u$ and $v$ in $G$ is called a {\it geodesic
path} if the length of the path $P$ is exactly $d(u,v)$. We call
$\sigma(v)=max_{u\in V}d(u,v)$ the {\it eccentricity} of $v\in V$.
In addition, $r(G)=min_{v\in V}\sigma(v)$ and $\rho(G)=max_{v\in
V}\sigma(v)$ are called the {\it radius} and {\it diameter} of $G$,
respectively. Without causing any confusion, we simply denote
$r(G),\rho(G)$ as $r,\rho$, respectively.
\end{definition}

A {\it path} graph is a simple graph whose vertices can be arranged
in a linear sequence in such a way that two vertices are adjacent if
they are consecutive in the sequence, and are nonadjacent otherwise.
Likewise, a {\it cycle} graph on three or more vertices is a simple
graph whose vertices can be arranged in a cyclic sequence in such a
way that two vertices are adjacent if they are consecutive in the
sequence, and are nonadjacent otherwise. Denote by $P_n$ and $C_n$
the path graph and the cycle graph on $n$ vertices, respectively.

A connected graph without any cycle is a \emph{tree}. A \emph{star}
of order $n$, denoted by $S_n$, is the tree with $n-1$ pendant
vertices. Its unique vertex with degree $n-1$ is called the
\emph{center vertex} of $S_n$. A simple connected graph is called
\emph{unicyclic} if it has exactly one cycle. We use $S^+_n$ to
denote the unicyclic graph obtained from the star $S_n$ by adding to
it an edge between two pendant vertices of $S_n$. Observe that a
tree and a unicyclic graph of order $n$ have exactly $n-1$ and $n$
edges, respectively. A \emph{bicyclic graph} is a graph of order $n$
with $n+1$ edges. A tree is called a \emph{double star} $S_{p,q}$ if
it is obtained from $S_{p+1}$ and $S_q$ by identifying a leaf of
$S_{p+1}$ with the center vertex of $S_q$. So, for the double star
$S_{p,q}$ with $n$ vertices, we have $p+q=n$. We call a double star
$S_{p,q}$ \emph{balanced} if $p=\lfloor\frac{n}{2}\rfloor$ and
$q=\lceil\frac{n}{2}\rceil$. A \emph{comet} is a tree composed of a
star and a pendant path. For any integers $n$ and $t$ with $2\leq
t\leq n-1$, we denote by $CS(n,t)$ the comet of order $n$ with $t$
pendant vertices, i.e., a tree formed by a path $P_{n-t}$ of which
one end vertex coincides with a pendant vertex of a star $S_{t+1}$
of order $t+1$.

\begin{definition}
The {\it $j$-sphere} of a vertex $v_i$ in $G=(V,E)\in \mathcal
{G}_{\cup C}$ is defined by the set
\begin{equation*}
S_j(v_i,G):=\{v\in V|d(v_i,v)=j,j\geq 1\}.
\end{equation*}
\end{definition}

\begin{definition}
Let $X$ be a discrete random variable by using alphabet $\mathcal
{A}$, and $p(x_i)=Pr(X=x_i)$ the probability mass function of $X$.
The {\it mean entropy} of $X$ is then defined by
\begin{equation*}
H(X):=\sum_{x_i\in\mathcal{A}}p(x_i)\log(p(x_i)).
\end{equation*}
\end{definition}

The concept of graph entropy introduced by Rashevsky in
\cite{Rashevsky} and Trucco in \cite{Trucco} is used to measure
structural complexity. Several graph invariants such as the number
of vertices, the vertex degree sequence, and extended degree
sequences have been used in the construction of graph entropy
measures. The main graph entropy measures can be divided into two
classes: classical measures and parametric measures. Classical
measures, denoted by $I(G,\tau)$, are defined relative to a
partition of a set $X$ of graph elements induced by an equivalence
relation $\tau$ on $X$. More precisely, let $X$ be a set of graph
elements (typically vertices), and let $\{X_i\}$, $1\leq i\leq k$,
be a partition of $X$ induced by $\tau$. Suppose further that
$p_i:=\frac{|X_i|}{|X|}$. Then
\begin{equation*}
I(G,\tau)=-\sum\limits_{i=1}^{k}p_i\log(p_i).
\end{equation*}
As mentioned in \cite{DM1}, Rashevsky \cite{Rashevsky} defined the
following graph entropy measure
\begin{equation}\label{eqAA}
I^V(G):=-\sum\limits_{i=1}^{k}\frac{|N_i|}{|V|}
\log\left(\frac{|N_i|}{|V|}\right)
\end{equation}
where $|N_i|$ denotes the number of topologically equivalent
vertices in the $i$-th vertex orbit of $G$ and $k$ is the number of
different orbits. Vertices are considered as topologically
equivalent if they belong to the same orbit of a graph. According to
\cite{DBVG}, we have that if a graph $G$ is vertex-transitive
\cite{MM,Harary}, then $I^V(G)=0$. Additionally, Trucco
\cite{Trucco} introduced a similar graph entropy measure
\begin{equation}\label{eqBB}
I^E(G):=-\sum\limits_{i=1}^{k}\frac{|N_i^E|}{|E|}
\log\left(\frac{|N_i^E|}{|E|}\right)
\end{equation}
where $|N^E_i|$ stands for the number of edges in the $i$-th edge
orbit of $G$. These two entropies are both classical measures, in
which special graph invariants (e.g., numbers of vertices, edges,
degrees, distances, etc.) and equivalence relations have given rise
to these measures of information contents. And thus far, a number of
specialized measures have been developed that are used primarily to
characterize the structural complexity of chemical graphs
\cite{Bonchev1,Bonchev-2,BR-1}.

In recent years, instead of inducing partitions and determining
their probabilities, researchers assign a probability value to each
individual element of a graph to derive graph entropy measures. This
leads to the other class of graph entropy measures: parametric
measures. Parametric measures are defined on graphs relative to
information functions. Such functions are not identically zero and
map graph elements (typically vertices) to nonnegative reals. Now we
give the precise definition for entropies belonging to parametric
measures.
\begin{definition}
Let $G\in \mathcal {G}_{\cup C}$ and let $S$ be a given set, e.g., a
set of vertices or paths, etc. Functions $f: S\rightarrow
\mathbb{R}_+$ play a role in defining information measures on graphs
and we call them information functions of $G$.
\end{definition}

\begin{definition}\label{defHH}
Let $f$ be an information function of $G$. Then
\begin{equation*}
p^f(v_i):=\frac{f(v_i)}{\sum\limits_{j=1}^{|V|}f(v_j)}.
\end{equation*}
Obviously,
\begin{equation*}
p^f(v_1)+p^f(v_2)+\cdots+p^f(v_n)=1, \ \text{where}\ n=|V|.
\end{equation*}
Hence, $(p^f(v_1),p^f(v_2),\cdots,p^f(v_n))$ forms a probability distribution.
\end{definition}

\begin{definition}\label{defGG}
Let $G$ be a finite graph and let $f$ be an information function of
$G$. Then
\begin{equation}\label{eqCC}
I_f(G):=-\sum\limits_{i=1}^{|V|}\frac{f(v_i)}{\sum_{j=1}^{|V|}f(v_j)}
\log \frac{f(v_i)}{\sum_{j=1}^{|V|}f(v_j)}
\end{equation}
\begin{equation}\label{eqDD}
I_f^\lambda(G):=\lambda\left(\log(|V|)+\sum\limits_{i=1}^{|V|}\frac{f(v_i)}{\sum_{j=1}^{|V|}f(v_j)}
\log \frac{f(v_i)}{\sum_{j=1}^{|V|}f(v_j)}\right)
\end{equation}
are families of information measures representing structural
information content of $G$, where $\lambda>0$ is a scaling constant.
$I_f$ is the entropy of $G$ which belongs to parametric measures and
$I_f^\lambda$ its information distance between maximum entropy and
$I_f$.
\end{definition}
The meaning of $I_f(G)$ and $I_f^\lambda(G)$ has been investigated
by calculating the information content of real and synthetic
chemical structures \cite{DM3}. Also, the information measures were
calculated using specific graph classes to study extremal values
and, hence, to detect the kind of structural information captured by
the measures.

In fact, there also exist graph entropy measures based on integral
though we do not focus on them in this paper. We introduce simply
here one such entropy: the tree entropy. For more details we refer
to \cite{Lyons1,Lyons2}. A graph $G=(V,E)$ with a distinguished
vertex $o$ is called a \emph{rooted graph}, which is denoted by
$(G,o)$ here. A \emph{rooted isomorphism} of rooted graphs is an
isomorphism of the underlying graphs that takes the root of one to
the root of the other. The \emph{simple random walk} on $G$ is the
Markov chain whose state space is $V$ and whose transition
probability from $x$ to $y$ equals the number of edges joining $x$
to $y$ divided by $d(x)$. The \emph{average degree} of $G$ is
$\frac{\sum_{x\in V}d(x)}{|V|}$.

Let $p_k(x;G)$ denote the probability that the simple random walk on
$G$ started at $x$ and back at $x$ after $k$ steps. Given a positive
integer $R$, a finite rooted graph $H$, and a probability
distribution $\rho$ on rooted graphs, let $p(R,H,\rho)$ denote the
probability that $H$ is rooted isomorphic to the ball of radius $R$
about the root of a graph chosen with distribution $\rho$. Define
the \emph{expected degree} of a probability measure $\rho$ on rooted
graphs to be
\begin{equation*}
\overline{d}(\rho):=\int d(o)d\rho(G,o).
\end{equation*}
For a finite graph $G$, let $U(G)$ denote the distribution of rooted
graphs obtained by choosing a uniform random vertex of $G$ as root
of $G$. Suppose that $\langle G_n\rangle$ is a sequence of finite
graphs and that $\rho$ is a probability measure on rooted infinite
graphs. We say that the \emph{random weak limit} of $\langle
G_n\rangle$ is $\rho$ if for any positive integer $R$, any finite
graph $H$, and any $\epsilon>0$, we have
$\lim\limits_{n\rightarrow\infty}
P[|p(R,H,U(G_n))-p(R,H,\rho)|>\epsilon]=0$.

Lyons \cite{Lyons1} proposed the tree entropy of a probability
measure $\rho$ on rooted infinite graphs
\begin{equation*}
\textbf{h}(\rho):=\int\left(\log d(o)-\sum_{k\geq 1}\frac{1}{k}p_k(o,G)\right)d\rho(G,o).
\end{equation*}
For labeled networks, i.e., labeled graphs, Lyons \cite{Lyons2} also
gave a definition of information measure, which is more general than
the tree entropy.
\begin{definition}\cite{Lyons2}
Let $\rho$ be a probability measure on rooted networks. We call $\rho$ unimodular if
\begin{equation*}
\int\sum_{x\in V(G)}f(G,o,x)d\rho(G,o)=\int\sum_{x\in V(G)}f(G,x,o)d\rho(G,o)
\end{equation*}
for all non-negative Borel functions $f$ on locally finite connected
networks with an ordered pair of distinguished vertices that is
invariant in the sense that for any (non-rooted) network isomorphism
$\gamma$ of $G$ and any $x,y\in V(G)$, we have $f(\gamma G,\gamma
x,\gamma y)=f(G,x,y)$.
\end{definition}

Actually, following the seminal paper of Shannon \cite{SW}, many
generalizations of the entropy measure have been proposed. An
important example of such a measure is called the {\it R\'{e}nyi
entropy} \cite{R} which is defined by
\begin{equation*}
I_\alpha^r(P):=\frac{1}{1-\alpha}
\log\left(\sum\limits_{i=1}^{n}(P_i)^\alpha\right),\ \alpha\neq1
\end{equation*}
where $n=|V|$ and $P:=(p_1,p_2,\cdots,p_{n})$. For further
discussion of the properties of R\'{e}nyi entropy, see \cite{Arndt}.
R\'{e}nyi and other general entropy functions allow for specifying
families of information measures that can be applied to graphs. Like
some generalized information measures that have been investigated in
information theory, Dehmer and Mowshowitz call these families
\emph{generalized graph entropies}. And in \cite{DM2}, they
introduced six distinct such entropies which are stated as follows.
\begin{definition}\label{defII}
Let $G=(V,E)$ be a graph on $n$ vertices. Then
\begin{eqnarray}
I^1_\alpha(G)&:=&\frac{1}{1-\alpha}\log\left(\sum\limits_{i=1}^{k}
\left(\frac{|X_i|}{|X|}\right)^\alpha\right)\label{eqII}\\
I^2_\alpha(G)_f&:=&\frac{1}{1-\alpha}\log\left(\sum\limits_{i=1}^{n}
\left(\frac{f(v_i)}{\sum_{j=1}^{n}f(v_j)}\right)^\alpha\right)\\
I^3_\alpha(G)&:=&\frac{\sum\limits_{i=1}^{k}\left(\frac{|X_i|}{|X|}
\right)^\alpha-1}{2^{1-\alpha}-1}\label{eqJJ}\\
I^4_\alpha(G)_f&:=&\frac{\sum\limits_{i=1}^{n}\left(\frac{f(v_i)}
{\sum_{j=1}^{n}f(v_j)}\right)^\alpha-1}{2^{1-\alpha}-1}\\
I^5(G)&:=&\sum\limits_{i=1}^{k}\frac{|X_i|}{|X|}
\left[1-\frac{|X_i|}{|X|}\right]\label{eqKK}\\
I^6(G)_f&:=&\sum\limits_{i=1}^{n}\frac{f(v_i)}{\sum_{j=1}^{n}f(v_j)}
\left[1-\frac{f(v_i)}{\sum_{j=1}^{n}f(v_j)}\right]
\end{eqnarray}
where $X$ is a set of graph elements (typically vertices), $\{X_i\}$
for $1\leq i\leq k$ is a partition of $X$ induced by the equivalence
relation $\tau$, $f$ is an information function of $G$ and
$\alpha\neq 1$.
\end{definition}

Parametric complexity measures have been proved useful in the study
of complexity associated with machine learning. And Dehmer et al.
\cite{DBVG2} showed that generalized graph entropies can be applied
to problems in machine learning such as graph classification and
clustering. Interestingly, these new generalized entropies have been
proved useful in demonstrating that hypotheses can be learned by
using appropriate data sets and parameter optimization techniques.

This chapter is organized as follows. Section 2 shows some
inequalities and extremal properties of graph entropies and
generalized graph entropies. Relationships between graph structures,
graph energies, topological indices and generalized graph entropies
are presented in Section 3, and the last section is a simple
summary.

\section{Inequalities and extremal properties on (generalized) graph entropies}
\label{ch1:sec2}

Thanks to the fact that graph entropy measures have been applied to
characterize the structures and complexities of graph-based systems
in various areas, identity and inequality relationships between
distinct graph entropies have been a hot and popular research topic.
In the meantime, extremal properties of graph entropies have also
been widely studied and lots of results were obtained.

\subsection{Inequalities for classical graph entropies and parametric measures}

Most of the graph entropy measures developed thus far have been
applied in mathematical chemistry and biology
\cite{Bonchev1,Bonchev2,DM1}. These measures have been used to
quantify the complexity of chemical and biological systems that can
be represented as graphs. Given the profusion of such measures, it
is useful to prove bounds for special graph classes or to study
interrelations among them. Dehmer et al. \cite{DME} gave
interrelation between the parametric entropy and a classical entropy
measure that is based on certain equivalence classes associated with
an arbitrary equivalence relation.
\begin{thm}\label{thmAA}\cite{DME}
Let $G=(V,E)$ be an arbitrary graph, and let $X_i,\ 1\leq i\leq k$,
be the equivalence classes associated with an arbitrary equivalence
relation on $X$. Suppose further that $f$ is an information function
with $f(v_i)>|X_i|$ for $1\leq i\leq k$ and
$c:=\frac{1}{\sum_{j=1}^{|V|}f(v_j)}$. Then
\begin{eqnarray*}
\frac{1}{|X|}I_f(G)&<&c\cdot I(G,\tau)-\sum\limits_{i=1}^{k}
\frac{|X_i|}{|X|}c\cdot\log(c)-\frac{\log(|X|)}{|X|}
\sum\limits_{i=1}^{k}p^f(v_i)\\
& &-\frac{1}{|X|}\sum\limits_{i=k+1}^{|V|}p^f(v_i)\log(p^f(v_i))
+\frac{1}{|X|}\sum\limits_{i=1}^{k}p^f(v_i)
\log\left(1+\frac{|X|}{c\cdot f(v_i)}\right)\\
& &+\sum\limits_{i=1}^{k}\log\left(\frac{p^f(v_i)}{|X|}+1\right).
\end{eqnarray*}
\end{thm}

Assume that $f(v_i)>|X_i|$, $1\leq i\leq k$, for some special graph
classes and take the set $X$ to be the vertex set $V$ of $G$. Three
corollaries of the above theorem on the upper bounds of $I_f(G)$ can
be obtained.
\begin{cor}\cite{DME}
Let $S_n$ be a star graph on $n$ vertices and suppose that $v_1$ is
the vertex with degree $n-1$. The remaining $n-1$ non-hub vertices
are labeled arbitrarily. $v_\mu$ stands for a non-hub vertex. Let
$f$ be an information function satisfying the conditions of Theorem
\ref{thmAA}. Let $V_1:=\{v_1\}$ and $V_2:=\{v_2,v_3,\cdots,v_n\}$
denote the orbits of the automorphism group of $S_n$ forming a
partition of $V$. Then
\begin{eqnarray*}
I_f(S_n)&<&p^f(v_1)\log\left(1+\frac{1}{p^f(v_1)}\right)
+p^f(v_\mu)\log\left(1+\frac{1}{p^f(v_\mu)}\right)\\
& &+\log(1+p^f(v_1))+\log(1+p^f(v_\mu))-\sum\limits_{i=2,i\neq\mu}^{n}
p^f(v_i)\log(p^f(v_i))\\
& &-(n-1)\cdot c\cdot\log[(n-1)c]-c\log(c).
\end{eqnarray*}
\end{cor}

\begin{cor}\cite{DME}
Let $G_n^I$ be an identity graph (a graph possessing a single graph
automorphism) on $n\geq 6$ vertices. $G_n^I$ has only the identity
automorphism and therefore each orbit is a singleton set, i,e.,
$|V_i|=1,\ 1\leq i\leq n$. Let $f$ be an information function
satisfying the conditions of Theorem \ref{thmAA}. Then
\begin{equation*}
I_f(G_n^I)<\sum\limits_{j=1}^{n}p^f(v_j)
\log\left(1+\frac{1}{p^f(v_j)}\right)
+\sum\limits_{j=1}^{n}\log(1+p^f(v_j))-n\cdot c\cdot\log(c).
\end{equation*}
\end{cor}

\begin{cor}\cite{DME}
Let $G_n^P$ be a path graph on $n$ vertices and let $f$ be an
information function satisfying the conditions of Theorem
\ref{thmAA}. If $n$ is even, $G_n^P$ posses $\frac{n}{2}$
equivalence classes $V_i$ and each $V_i$ contains $2$ vertices. Then
\begin{eqnarray*}
I_f(G_n^P)&<&\sum\limits_{j=1}^{\frac{n}{2}}p^f(v_j)
\log\left(1+\frac{1}{p^f(v_j)}\right)
+\sum\limits_{j=1}^{\frac{n}{2}}\log(1+p^f(v_j))\\
& &-\sum\limits_{j=\frac{n}{2}+1}^{n}p^f(v_j)\log(1+p^f(v_j))-n\cdot c\cdot\log(2c).
\end{eqnarray*}
If $n$ is odd, then there exist $n-\lfloor\frac{n}{2}\rfloor$
equivalence classes, $n-\lfloor\frac{n}{2}\rfloor-1$ that have $2$
elements and only one class containing a single element. This
implies that
\begin{eqnarray*}
I_f(G_n^P)&<&\sum\limits_{j=1}^{n-\lfloor\frac{n}{2}\rfloor}p^f(v_j)
\log\left(1+\frac{1}{p^f(v_j)}\right)
+\sum\limits_{j=1}^{n-\lfloor\frac{n}{2}\rfloor}\log(1+p^f(v_j))\\
& &-\sum\limits_{j=n-\lfloor\frac{n}{2}\rfloor+1}^{n}
p^f(v_j)\log(p^f(v_j))-(n-\lfloor\frac{n}{2}\rfloor-1)\cdot 2c\cdot\log(2c)\\
& &-c\cdot\log(c).
\end{eqnarray*}
\end{cor}

Assuming different initial conditions, Dehmer et al. \cite{DME}
derived additional inequalities between classical and parametric
measures.
\begin{thm}\cite{DME}
Let $G$ be an arbitrary graph and $p^f(v_i)<|X_i|$. Then
\begin{eqnarray*}
\frac{1}{|X|}I_f(G)&>&I(G,\tau)-\frac{1}{|X|}
\sum\limits_{i=k+1}^{|V|}p^f(v_i)\log(p^f(v_i))
-\frac{\log(|X|)}{|X|}\sum\limits_{i=1}^{k}p^f(v_i)\\
& &-\frac{1}{|X|}\sum\limits_{i=1}^{k}|X_i|
\log\left(1+\frac{|X|}{|X_i|}\right)-\sum\limits_{i=1}^{k}
\log\left(1+\frac{|X_i|}{|X|}\right).
\end{eqnarray*}
\end{thm}

\begin{thm}\cite{DME}
Let $G$ be an arbitrary graph with $p_i$ being the probabilities
such that $p_i<f(v_i)$. Then
\begin{eqnarray*}
\frac{1}{c}I(G,\tau)&>&I_f(G)+\frac{\log(c)}{c}
+\sum\limits_{i=k+1}^{|V|}p^f(v_i)\log(p^f(v_i))\\
& &-\sum\limits_{i=1}^{k}\log(p^f(v_i))
-\sum\limits_{i=1}^{k}
\log\left(1+\frac{1}{p^f(v_i)}\right)(1+p^f(v_i)).
\end{eqnarray*}
\end{thm}

For identity graphs, they also obtained a general upper bound for
the parametric entropy measure.

\begin{cor}\cite{DME}
Let $G_n^I$ be an identity graph on $n$ vertices. Then
\begin{eqnarray*}
I_f(G_n^I)&<&\log(n)-c\cdot \log(c)+\sum\limits_{i=1}^{n}\log(p^f(v_i))\\
& &+\sum\limits_{i=1}^{n}
\log\left(1+\frac{1}{p^f(v_i)}\right)(1+p^f(v_i)).
\end{eqnarray*}
\end{cor}

\subsection{Graph entropy inequalities with information functions $f^V$, $f^P$ and $f^C$}

In complex networks, information-theoretical methods are important
for analyzing and understanding information processing. One major
problem is to quantify structural information in networks based on
so-called information functions. Considering a complex network as
an undirected connected graph and based on such information
functions, one can directly obtain different graphs entropies.

Now we define two information functions $f^V(v_i),\ f^P(v_i)$,
based on metrical properties of graphs, and a novel information
function $f^C(v_i)$, based on a vertex centrality measure.
\begin{definition}\cite{Dehmer}\label{defAA}
Let $G=(V,E)\in \mathcal {G}_{\cup C}$. For a vertex $v_i\in V$, we
define the information function
\begin{equation*}
f^V(v_i):=\alpha^{c_1|S_1(v_i,G)|+c_2|S_2(v_i,G)|+\cdots
+c_\rho|S_\rho(v_i,G)|}, \quad c_k>0,\ 1\leq k\leq \rho,\ \alpha>0,
\end{equation*}
where the $c_k$ are arbitrary real positive coefficients,
$S_j(v_i,G)$ denotes the $j$-sphere of $v_i$ regarding $G$ and
$|S_j(v_i,G)|$ its cardinality, respectively.
\end{definition}

Before giving the definition of the information function
$f^P(v_i)$, we introduce the following concepts first.
\begin{definition}\cite{Dehmer}\label{defKK}
Let $G=(V,E)\in \mathcal {G}_{UC}$. For a vertex $v_i\in V$ we
determine the set $S_j(v_i,G)=\{v_{u_j},v_{w_j},\cdots,v_{x_j}\}$
and define associated paths
\begin{eqnarray*}
P^j_1(v_i)&=&(v_i,v_{u_1},v_{u_2},\cdots,v_{u_j}),\\
P^j_2(v_i)&=&(v_i,v_{w_1},v_{w_2},\cdots,v_{w_j}),\\
&\vdots&\\
P^j_{k_j}(v_i)&=&(v_i,v_{x_1},v_{x_2},\cdots,v_{x_j}),
\end{eqnarray*}
and their edge sets
\begin{eqnarray*}
E_1&=&\{\{v_i,v_{u_1}\},\{v_{u_2},v_{u_3}\},\cdots,\{v_{u_{j-1}},v_{u_j}\}\},\\
E_2&=&\{\{v_i,v_{w_1}\},\{v_{w_2},v_{w_3}\},\cdots,\{v_{w_{j-1}},v_{w_j}\}\},\\
&\vdots&\\
E_{k_j}&=&\{\{v_i,v_{x_1}\},\{v_{x_2},v_{x_3}\},\cdots,
\{v_{x_{j-1}},v_{x_j}\}\}.
\end{eqnarray*}
Now we define the graph
$\mathscr{L}_G(v_i,j)=(V_{\mathscr{L}},E_{\mathscr{L}})\subseteq G$
as the local information graph regarding $v_i\in V$ with respect to
$f$, where
\begin{equation*}
V_{\mathscr{L}}:=\{v_i,v_{u_1},v_{u_2},\cdots,v_{u_j}\}\cup
\{v_i,v_{w_1},v_{w_2},\cdots,v_{w_j}\}\cup \cdots\cup
\{v_i,v_{x_1},v_{x_2},\cdots,v_{x_j}\}
\end{equation*}
and
\begin{equation*}
E_{\mathscr{L}}:=E_1\cup E_2\cup \cdots\cup E_{k_j}.
\end{equation*}
Further, $j=j(v_i)$ is called the local information radius regarding $v_i$.
\end{definition}

\begin{definition}\cite{Dehmer}\label{defBB}
Let $G=(V,E)\in \mathcal {G}_{\cup C}$. For each vertex $v_i\in V$
and for $j\in 1,2, \cdots, \rho$, we determine the local information
graph $\mathscr{L}_G(v_i,j)$ where $\mathscr{L}_G(v_i,j)$ is induced
by the paths $P_1^j(v_i), P_2^j(v_i), \cdots, P_{k_j}^j(v_i)$. The
quantity $l(P_{\mu}^{j}(v_i))\in \mathbb{N},\
\mu\in\{1,2,\cdots,k_j\}$ denotes the length of $P_{\mu}^{j}(v_i)$
and
\begin{equation*}
l(P(\mathscr{L}_G(v_i,j))):=\sum\limits_{\mu=1}^{k_j}l(P_{\mu}^{j}(v_i))
\end{equation*}
expresses the sum of the path lengths associated to each
$\mathscr{L}_G(v_i,j)$. Now we define the information function
$f^P(v_i)$ as
\begin{equation*}
f^P(v_i):=\alpha^{b_1l(P(\mathscr{L}_G(v_i,1)))+b_2l(P(\mathscr{L}_G(v_i,2)))
+\cdots+b_\rho l(P(\mathscr{L}_G(v_i,\rho)))}
\end{equation*}
where $b_k>0,\ 1\leq k\leq \rho,\ \alpha>0$ and $b_k$ are arbitrary
real positive coefficients.
\end{definition}

\begin{definition}\cite{Dehmer}\label{defCC}
Let $G=(V,E)\in \mathcal {G}_{\cup C}$ and $\mathscr{L}_G(v_i,j)$
denote the local information graph defined as above for each vertex
$v_i\in V$. We define $f^C(v_i)$ as
\begin{equation*}
f^C(v_i):=\alpha^{a_1\beta^{\mathscr{L}_G(v_i,1)}(v_i)
+a_2\beta^{\mathscr{L}_G(v_i,2)}(v_i)+\cdots
+a_\rho\beta^{\mathscr{L}_G(v_i,\rho)}(v_i)}
\end{equation*}
where $\beta\leq1,\ a_k>0,\ 1\leq k\leq \rho,\ \alpha>0$, $\beta$ is
a certain vertex centrality measure, $\beta^{\mathscr{L}_G(v_i,j)}(v_i)$ expresses that we apply $\beta$ to $v_i$ regarding $\mathscr{L}_G(v_i,j)$ and $a_k$ are arbitrary real
positive coefficients.
\end{definition}

By applying Definitions \ref{defAA}, \ref{defBB}, \ref{defCC} and
Equation \ref{eqCC}, we obviously obtain the following three special
graph entropies:
\begin{equation}\label{eqEE}
I_{f^V}(G)=-\sum\limits_{i=1}^{|V|}\frac{f^V(v_i)}{\sum_{j=1}^{|V|}f^V(v_j)}
\log \frac{f^V(v_i)}{\sum_{j=1}^{|V|}f^V(v_j)}
\end{equation}
\begin{equation}\label{eqFF}
I_{f^P}(G)=-\sum\limits_{i=1}^{|V|}\frac{f^P(v_i)}{\sum_{j=1}^{|V|}f^P(v_j)}
\log \frac{f^P(v_i)}{\sum_{j=1}^{|V|}f^P(v_j)}
\end{equation}
and
\begin{equation}
I_{f^C}(G)=-\sum\limits_{i=1}^{|V|}\frac{f^C(v_i)}{\sum_{j=1}^{|V|}f^C(v_j)}
\log \frac{f^C(v_i)}{\sum_{j=1}^{|V|}f^C(v_j)}.
\end{equation}

The entropy measures based on the defined information
functions ($f^V$, $f^P$ and $f^C$) can detect the structural
complexity between graphs and therefore capture important structural
information meaningfully. In \cite{Dehmer}, Dehmer investigated
relationships between the above graph entropies and analyzed the
computational complexity of these entropy measures.
\begin{thm}\cite{Dehmer}
Let $G=(V,E)\in \mathcal {G}_{\cup C}$ and let $f^V$, $f^P$ and
$f^C$ be information functions defined above. For the associated
graph entropies, it holds the inequality
\begin{equation*}
I_{f^V}(G)>\alpha^{\rho[\phi^P\omega^P-\varphi]}
\left[I_{f^P}(G)-log\left(\alpha^{\rho[\phi^P\omega^P-\varphi]}\right)\right],
\quad \alpha>1
\end{equation*}
where $\omega^P=max_{1\leq i\leq |V|}\omega^P(v_i)$,
$\omega^P(v_i)=max_{1\leq j\leq \rho}l(P(\mathscr{L}_G(v_i,j)))$,
$\phi^P=max_{1\leq j\leq \rho}b_j$ and $\varphi=min_{1\leq j\leq
\rho}c_j$; and
\begin{equation*}
I_{f^V}(G)<\alpha^{\rho[\varphi^Cm^C-\phi\omega]}
\left[I_{f^C}(G)-log\left(\alpha^{\rho[\varphi^Cm^C-\phi\omega]}\right)\right],
\quad \alpha>1
\end{equation*}
where $\phi=max_{1\leq j\leq \rho}c_j$, $\varphi^C=min_{1\leq j\leq
\rho}a_j$, $m^C=min_{1\leq i\leq |V|}m^C(v_i)$, $\omega=max_{1\leq
i\leq |V|}(\omega(v_i))$ and $\omega(v_i)=max_{1\leq j\leq
\rho}|S_j(v_i,G)|$.
\end{thm}

\begin{thm}\cite{Dehmer}
The time complexity to compute the entropies $I_{f^V}(G)$,
$I_{f^P}(G)$ and $I_{f^C}(G)$ for $G\in \mathcal {G}_{\cup C}$ is
$O(|V|^3)$.
\end{thm}

\subsection{Information theoretic measures of UHG graphs}

Let $G$ be an undirected graph with vertex set $V$, edge set $E$ and
$N=|V|$ vertices. We call the function $L:\ V\rightarrow
\mathscr{L}$ \emph{multi-level function}, which assigns to all vertices of $G$ an element $l\in \mathscr{L}$ that corresponds to the level it
will be assigned. Then a \emph{universal hierarchical graph} is defined by a vertex set $V$, an edge set $E$, a level set $\mathscr{L}$ and a
multi-level function $L$. The vertex and edge sets define the
connectivity and the level set and the multi-level function induce a
hierarchy between the vertices of $G$. We denote the class of
universal hierarchical graphs (UHG) by $\mathscr{G}_{UH}$.

Rashevsky \cite{Rashevsky} suggested to partition a graph and to
assign probabilities $p_i$ to all partitions in a certain way. Here,
for a graph $G=(V,E)\in \mathscr{G}_{UH}$, such a partition is given
naturally by the hierarchical levels of $G$. This property directly
leads to the definition of its graph entropies.
\begin{definition}
We assign a discrete probability distribution $P^n$ to a graph $G\in
\mathscr{G}_{UH}$ with $\mathscr{L}$ in the following way: $P^n:\
\mathscr{L}\rightarrow [0,1]^{|\mathscr{L}|}$ with
$p_i^n:=\frac{n_i}{N}$, where $n_i$ is the number of vertices on
level $i$. The vertex entropy of $G$ is defined as
\begin{equation*}
H^n(G)=-\sum\limits_{i}^{|\mathscr{L}|}p_i^n\log(p_i^n).
\end{equation*}
\end{definition}

\begin{definition}\label{defJJ}
We assign a discrete probability distribution $P^e$ to a graph $G\in
\mathscr{G}_{UH}$ with $\mathscr{L}$ in the following way: $P^e:\
\mathscr{L}\rightarrow [0,1]^{|\mathscr{L}|}$ with
$p_i^e:=\frac{e_i}{E^0}$. Here $e_i$ is the number of edges incident
with the vertices on level $i$ and $E^0=2|E|$. The edge entropy of
$G$ is defined as
\begin{equation*}
H^e(G)=-\sum\limits_{i}^{|\mathscr{L}|}p_i^e\log(p_i^e).
\end{equation*}
\end{definition}

Emmert-Streib and Dehmer \cite{ED} focused on the extremal
properties of entropy measures of UHG graphs. In addition, they
proposed the concept of joint entropy of universal hierarchical
graphs and further studied its extremal properties.
\begin{thm}\cite{ED}
For $G\in \mathscr{G}_{UH}$ with $N$ vertices and $|\mathscr{L}|$
levels. The condition for $G$ to have maximum vertex entropy is

(1) \ if $\frac{N}{|\mathscr{L}|}\in \mathbb{N}:\ p_i=\frac{n}{N}$
with $n=\frac{N}{|\mathscr{L}|}$, or

(2) \ if $\frac{N}{|\mathscr{L}|}\in \mathbb{R}:$
\begin{equation*}
p_i=\left\{
\begin{array}{ll}
\frac{n}{N}:\ &1\leq i\leq I_1,\\
\frac{n-1}{N}:\ &I_1+1\leq i\leq |\mathscr{L}|=I_1+I_2.
\end{array}
\right.
\end{equation*}
\end{thm}

\begin{thm}\cite{ED}
For $G\in \mathscr{G}_{UH}$ with $|E|$ edges and $|\mathscr{L}|$
levels. The condition for $G$ to have maximum edge entropy is

(1) \ if $\frac{E^0}{|\mathscr{L}|}\in \mathbb{N}:\
p_i=\frac{e}{E^0}$ with $e=\frac{E^0}{|\mathscr{L}|}$, or

(2) \ if $\frac{E^0}{|\mathscr{L}|}\in \mathbb{R}:$
\begin{equation*}
p_i=\left\{
\begin{array}{ll}
\frac{e}{E^0}:\ &1\leq i\leq I_1,\\
\frac{e-1}{E^0}:\ &I_1+1\leq i\leq |\mathscr{L}|=I_1+I_2.
\end{array}
\right.
\end{equation*}
\end{thm}

Now we give two joint probability distributions on $G\in
\mathscr{G}_{UH}$ and introduce two joint entropies for $G$.
\begin{definition}\label{defDD}
A discrete joint probability distribution on $G\in \mathscr{G}_{UH}$
is naturally given by $p_{ij}:=p_i^np_j^e$. The resulting joint
entropy of $G$ is given by
\begin{equation*}
H_2(G)=-\sum\limits_{i}^{|\mathscr{L}|}\sum\limits_{j}^{|\mathscr{L}|}
p_{ij}\log(p_{ij}).
\end{equation*}
\end{definition}

\begin{definition}\label{defEE}
A discrete joint probability distribution on $G\in \mathscr{G}_{UH}$
can also be given by
\begin{equation*}
p_{ij}=\left\{
\begin{array}{ll}
\frac{p_i^np_i^e}{\sum_jp_j^np_j^e}:\ &i=j,\\
0:\ &i\neq j.
\end{array}
\right.
\end{equation*}
The resulting joint entropy of $G$ is given by
\begin{equation*}
H'_2(G)=-\sum\limits_{i}^{|\mathscr{L}|}\sum\limits_{j}^{|\mathscr{L}|}
p_{ij}\log(p_{ij}).
\end{equation*}
\end{definition}

Interestingly, the extremal property of joint entropy in Definition
\ref{defEE} for $\frac{N}{|\mathscr{L}|}\in \mathbb{N}$ or
$\frac{E^0}{|\mathscr{L}|}\in \mathbb{N}$ is similar to that of
joint entropy in Definition \ref{defDD}.
\begin{thm}\cite{ED}
For $G\in \mathscr{G}_{UH}$ with $N$ vertices, $|E|$ edges and
$|\mathscr{L}|$ levels. The condition for $G$ to have maximum joint
entropy is

(1) \ if $\frac{N}{|\mathscr{L}|}\in \mathbb{N}$ and
$\frac{E^0}{|\mathscr{L}|}\in \mathbb{N}$: $p_i^n=\frac{n}{N}$ with
$n=\frac{N}{|\mathscr{L}|}$ and $p_i^e=\frac{e}{E^0}$ with
$e=\frac{E^0}{|\mathscr{L}|}$

(2) \ if $\frac{N}{|\mathscr{L}|}\in \mathbb{N}$ and
$\frac{E^0}{|\mathscr{L}|}\in \mathbb{R}$: $p_i^n=\frac{n}{N}$ with
$n=\frac{N}{|\mathscr{L}|}$ and
\begin{equation*}
p_{i}^e=\left\{
\begin{array}{ll}
\frac{e}{E^0}:\ &1\leq i\leq I_1^e,\\
\frac{e-1}{E^0}:\ &I_1^e+1\leq i\leq L=|\mathscr{L}|=I_1^e+I_2^e
\end{array}
\right.
\end{equation*}

(3) \ if $\frac{N}{|\mathscr{L}|}\in \mathbb{R}$ and
$\frac{E^0}{|\mathscr{L}|}\in \mathbb{N}$: $p_i^e=\frac{e}{E^0}$
with $e=\frac{E^0}{|\mathscr{L}|}$ and
\begin{equation*}
p_{i}^n=\left\{
\begin{array}{ll}
\frac{n}{N}:\ &1\leq i\leq I_1^n,\\
\frac{n-1}{N}:\ &I_1^n+1\leq i\leq |\mathscr{L}|=I_1^n+I_2^n
\end{array}
\right.
\end{equation*}

(4) \ if $\frac{N}{|\mathscr{L}|}\in \mathbb{R}$ and
$\frac{E^0}{|\mathscr{L}|}\in \mathbb{R}$:
\begin{equation*}
p_{i}^e=\left\{
\begin{array}{ll}
\frac{e}{E^0}:\ &1\leq i\leq I_1^e,\\
\frac{e-1}{E^0}:\ &I_1^e+1\leq i\leq |\mathscr{L}|=I_1^e+I_2^e,
\end{array}
\right.
\end{equation*}
\begin{equation*}
p_{i}^n=\left\{
\begin{array}{ll}
\frac{n}{N}:\ &1\leq i\leq I_1^n,\\
\frac{n-1}{N}:\ &I_1^n+1\leq i\leq |\mathscr{L}|=I_1^n+I_2^n.
\end{array}
\right.
\end{equation*}
\end{thm}

Note that the algorithmic computation of information-theoretical
measures always requires polynomial time complexity. Also in
\cite{ED}, Emmert-Streib and Dehmer provided some results about the
time complexity to compute the vertex and edge entropy introduced as
above.
\begin{thm}\cite{ED}
The time complexity to compute the vertex entropy (or edge entropy,
which is defined in Definition \ref{defJJ}) of an UHG graph $G$ with
$N$ vertices and $|\mathscr{L}|$ hierarchical levels is $O(N)(or\
O(N^2))$.
\end{thm}

Let $e_{li}$ denote the number of edges the $i$-th vertex has on
level $l$ and $\pi_l(\cdot)$ be a permutation function on level $l$
that orders the $e_{li}$'s such that $e_{li}\geq e_{l,i+1}$ with
$i=\pi_l(k)$ and $i+1=\pi_l(m)$. This leads to an $L\times N_{max}$
matrix $M$ whose elements correspond to $e_{li}$ where $i$ is the
column index and $l$ the row index. The number $N_{max}$ is the
maximal number of vertices a level can have. Additionally, the
authors \cite{ED} also introduced another edge entropy and studied
the time complexity to compute it, which we will state it in the
following.
\begin{definition}\label{defFF}
We assign a discrete probability distribution $P^e$ to a graph $G\in
\mathscr{G}_{UH}$ with $\mathscr{L}$ in the following way: $P^e:\
\mathscr{L}\rightarrow [0,1]^{|\mathscr{L}|}$ with
$p_i^e:=\frac{1}{N_{max}}\sum_i\frac{e_{li}}{M_i}$,
$M_i=\sum_ie_{li}$. The edge entropy of $G$ is now defined as
\begin{equation*}
H^e(G)=-\sum\limits_{i}^{|\mathscr{L}|}p_i^e\log(p_i^e).
\end{equation*}
\end{definition}

\begin{thm}\cite{ED}
The time complexity to compute the edge entropy in Definition
\ref{defFF} of an UHG graph $G$ with $N$ vertices and
$|\mathscr{L}|$ hierarchical levels is $O(|\mathscr{L}|\cdot
max((N^0)^2,(N^1)^2,\cdots,(N^{|\mathscr{L}|})^2))$. Here, $N^l$
with $l\in\{0,\cdots,|\mathscr{L}|\}$ is the number of vertices on
level $l$.
\end{thm}

\subsection{Bounds for the entropies of rooted trees and generalized trees}\label{subsAA}

The investigation of topological aspects of chemical structures
constitutes a major part of the research in chemical graph theory
and mathematical chemistry \cite{BR1,DGJ,GP,T}. There is a universe
of problems dealing with trees for modeling and analyzing chemical
structures. However, also rooted trees have wide applications in
chemical graph theory such as enumeration and coding problems of
chemical structures and so on.

Here, a \emph{hierarchical graph} means a graph having a distinct
vertex that is called a root and we also call it a \emph{rooted
graph}. Dehmer et al. \cite{DBE} derived bounds for the entropies of
hierarchical graphs in which they chose the classes of rooted trees
and so-called generalized trees. To start with the results of
entropy bounds, we first define the graph classes mentioned above.
\begin{definition}
An undirected graph is called undirected tree if this graph is
connected and cycle free. An undirected rooted tree $\mathcal
{T}=(V,E)$ is an undirected graph which has exactly one vertex $r\in
V$ for which every edge is directed away from the root $r$. Then,
all vertices in $\mathcal {T}$ are uniquely accessible from $r$. The
level of a vertex $v$ in a rooted tree $\mathcal {T}$ is simply the
length of the path from $r$ to $v$. The path with the largest path
length from the root to a leaf is denoted as $h$.
\end{definition}

\begin{definition}
As a special case of $\mathcal {T}=(V,E)$ we also define an ordinary
$w$-tree denoted as $\mathcal {T}_w$ where $w$ is a natural number.
For the root vertex $r$, it holds $d(r)=w$ and for all internal
vertices $r\in V$ holds $d(v)=w+1$. Leaves are vertices without
successors. A $w$-tree is fully occupied, denoted by $\mathcal
{T}_w^o$ if all leaves possess the same height $h$.
\end{definition}

\begin{definition}
Let $\mathcal {T}=(V,E_1)$ be an undirected finite rooted tree.
$|L|$ denotes the cardinality of the level set
$L:=\{l_0,l_1,\cdots,l_h\}$. The longest length of a path in
$\mathcal {T}$ is denoted as $h$. It holds $h=|L|-1$. The mapping $\Lambda:V\rightarrow L$ is surjective and it is called a
multi level function if it assigns to each vertex an element of the
level set $L$. A graph $H=(V,E_{G\mathcal {T}})$ is called a finite,
undirected generalized tree if its edge set can be represented by
the union $E_{G\mathcal {T}}: = E_1\cup E_2\cup E_3$, where

$\bullet$ $E_1$ forms the edge set of the underlying undirected
rooted tree $\mathcal {T}$.

$\bullet$ $E_2$ denotes the set of horizontal across-edges, i.e., an
edge whose incident vertices are at the same level $i$.

$\bullet$ $E_3$ denotes the set of edges whose incident vertices are
at different levels.
\end{definition}

Note that the definition of graph entropy here are the same as
Definition \ref{defAA} and Equation \ref{eqEE}. Inspired by the
technical assertion proved in \cite{ED}, Dehmer et al. \cite{DBE}
studied bounds for the entropies of rooted trees and so-called
generalized trees. Here we give the entropy bounds of rooted trees
first.
\begin{thm}\cite{DBE}
Let $\mathcal{T}$ be a rooted tree. For the entropy of
$\mathcal{T}$, it holds the inequality
\begin{equation*}
I_{f^V}(\mathcal{T})>\alpha^{\rho[\phi\cdot\omega-\varphi]}
\left[I_g(\mathcal{T})-
\log\left(\alpha^{\rho[\phi\cdot\omega-\varphi]}\right)\right],\
\forall \alpha>1
\end{equation*}
where
\begin{equation*}
I_{g}(\mathcal{T}):=-\left[g(v_{01})+\sum\limits_{i=1}^{h}
\sum\limits_{k=1}^{\sigma_i}(v_{ik})\log(g(v_{ik}))\right]
\end{equation*}
and $\omega:=max_{0\leq i\leq h, 1\leq k\leq
\sigma_i}\omega(v_{ik})$, $\omega(v_{ik}):=max_{1\leq j\leq
\rho}|S_j(v_{ik},\mathcal{T})|$, $\phi:=max_{1\leq j\leq \rho}c_j$,
$\varphi:=min_{1\leq j\leq \rho}c_j$, $v_{ik}$ denotes the $k$-th
vertex on the $i$-th level, $1\leq i\leq h$, $1\leq k\leq\sigma_i$
and $\sigma_i$ denotes the number of vertices on level $i$.
\end{thm}

As directed corollaries, special bounds for the corresponding
entropies have been obtained by considering special classes of
rooted trees.
\begin{cor}\cite{DBE}
Let $\mathcal{T}_w^o$ be a fully occupied $w$-tree. For the entropy
of $\mathcal{T}_w^o$ holds
\begin{equation*}
I_{f^V}(\mathcal{T}^o)>\alpha^{2h[\phi\cdot\omega^h-\varphi]}
\left[I_g(\mathcal{T}_w^o)-
\log\left(\alpha^{2h[\phi\cdot\omega^h-\varphi]}\right)\right],\
\forall \alpha>1.
\end{equation*}
\end{cor}

\begin{cor}\cite{DBE}
Let $\mathcal{T}_w$ be an ordinary $w$-tree. For the entropy of
$\mathcal{T}_w$ holds
\begin{equation*}
I_{f^V}(\mathcal{T})>\alpha^{\rho[\phi\cdot\omega^h-\varphi]}
\left[I_g(\mathcal{T})-
\log\left(\alpha^{\rho[\phi\cdot\omega^h-\varphi]}\right)\right],\
\forall \alpha>1.
\end{equation*}
\end{cor}

Next we will state the entropy bounds for generalized trees. In
fact, the entropy of a specific generalized tree can be
characterized by the entropy of another generalized tree that is
extremal with respect to a certain structural property; see the
following theorems.
\begin{thm}\cite{DBE}
Let $H=(V,E_{G\mathcal{T}})$ be a generalized tree with
$E_{G\mathcal{T}}:=E_1\cup E_2$, i.e., $H$ possesses across-edges
only. Starting from $H$, we define $H^*$ as the generalized tree
with the maximal number of across-edges on each level $i$, $1\leq
i\leq h$.

$\bullet$ First, there exist positive real coefficients $c_k$ which
satisfy the inequality system
\begin{equation*}
c_1|S_1(v_{ik},H^*)|+c_2|S_2(v_{ik},H^*)|+\cdots
+c_\rho|S_\rho(v_{ik},H^*)|
\end{equation*}
\begin{equation*}
>c_1|S_1(v_{ik},H)|+c_2|S_2(v_{ik},H)|+\cdots
+c_\rho|S_\rho(v_{ik},H)|
\end{equation*}
where $0\leq i\leq h$, $1\leq k\leq \sigma_i$, $c_j\geq 0$, $1\leq
j\leq \rho$ and $\sigma_i$ denotes the number of vertices on level
$i$.

$\bullet$ Second, it holds
\begin{equation*}
I_{f^V}(H)>\alpha^{\rho[\phi^*\cdot\omega^*-\varphi]}
\left[I_{f^V}(H^*)-
\log\left(\alpha^{\rho[\phi^*\cdot\omega^*-\varphi]}\right)\right],\ \forall \alpha>1.
\end{equation*}
\end{thm}

\begin{thm}\cite{DBE}
Let $H=(V^H,E)$ be an arbitrary generalized tree and let
$H_{|V|,|V|}$ be the complete generalized tree such that $|V^H|\leq
|V|$. It holds
\begin{equation*}
I_{f^V}(H)\leq I_{f^V}(H_{|V|,|V|}).
\end{equation*}
\end{thm}

\subsection{Information inequalities for $I_f(G)$ based on
different information functions}\label{subsBB}

We begin this section with some definition and notation.
\begin{definition}
Parameterized exponential information function using $j$-spheres:
\begin{equation}
f_{P}(v_i)=\beta^{\sum\limits_{j=1}^{\rho(G)}c_j|S_j(v_i,G)|}
\label{eqGG}
\end{equation}
where $\beta>0$ and $c_k>0$ for $1\leq k\leq \rho(G)$.
\end{definition}

\begin{definition}
Parameterized linear information function using $j$-spheres:
\begin{equation}
f_{P'}(v_i)=\sum\limits_{j=1}^{\rho(G)}c_j|S_j(v_i,G)|\label{eqHH}
\end{equation}
where $c_k>0$ for $1\leq k\leq \rho(G)$.
\end{definition}

Let $L_G(v,j)$ be the subgraph induced by the shortest path starting
from the vertex $v$ to all the vertices at distance $j$ in $G$.
Then, $L_G(v,j)$ is called the local information graph regarding $v$
with respect to $j$, which is defined as in Definition \ref{defKK}
\cite{Dehmer}. A local centrality measure that can be applied to
determine the structural information content of a network
\cite{Dehmer} is then defined as follows. We assume that $G=(V,E)$
is a connected graph with $|V|=n$ vertices.
\begin{definition}
The closeness centrality of the local information graph is defined by
\begin{equation*}
\gamma(v;L_G(v,j))=\frac{1}{\sum\limits_{x\in L_G(v,j)}d(v,x)}.
\end{equation*}
\end{definition}

Similar to the $j$-sphere functions, we define further functions
based on the local centrality measure as follows.
\begin{definition}
Parameterized exponential information function using local centrality measure:
\begin{equation*}
f_C(v_i)=\alpha^{\sum_{j=1}^{n}c_j\gamma(v_i;L_G(v_i,j))},
\end{equation*}
where $\alpha>0$, $c_k>0$ for $1\leq k\leq \rho(G)$.
\end{definition}

\begin{definition}
Parameterized linear information function using local centrality measure:
\begin{equation*}
f_{C'}(v_i)=\sum\limits_{j=1}^{n}c_j\gamma(v_i;L_G(v_i,j)),
\end{equation*}
where $c_k>0$ for $1\leq k\leq \rho(G)$.
\end{definition}
Recall that entropy measures have been used to quantify the
information content of the underlying networks and functions
became more meaningful when we choose the coefficients to emphasize
certain structural characteristics of the underlying graphs.

Now, we first present closed form expressions for the graph entropy $I_f(S_n)$.
\begin{thm}\cite{DS}
Let $S_n$ be a star graph on $n$ vertices. Let
$f\in\{f_P,f_{P'},f_C,f_{C'}\}$ be the information functions as defined
above. The graph entropy is given by
\begin{equation*}
I_f(S_n)=-x\log_2x-(1-x)\log_2\left(\frac{1-x}{n-1}\right),
\end{equation*}
where $x$ is the probability of the central vertex of $S_n$:
\begin{eqnarray*}
x&=&\frac{1}{1+(n-1)\beta^{(c_2-c_1)(n-2)}},\quad \text{if}\ f=f_P,\\
x&=&\frac{c_1}{2c_1+c_2(n-2)},\quad \text{if}\ f=f_{P'},\\
x&=&\frac{1}{1+(n-1)\alpha^{c_1\left(\frac{n-2}{n-1}\right)
+c_2\left(\frac{1}{2n-3}\right)}},\quad \text{if}\ f=f_C,\\
x&=&\frac{c_1}{c_1(1+(n-1)^2)
+c_2\left(\frac{(n-1)^2}{2n-3}\right)},\quad \text{if}\ f=f_{C'}.
\end{eqnarray*}
\end{thm}

Note that to compute a closed form expression even for a path is not
always simple. To illustrate this, we present the graph entropy
$I_{f_{P'}}(P_n)$ by choosing particular values for its
coefficients.
\begin{thm}\cite{DS}
Let $P_n$ be a path graph and set $c_1:=\rho(P_n)=n-1,\
c_2:=\rho(P_n)-1=n-2,\cdots,c_\rho:=1$. We have
\begin{eqnarray*}
I_{f_{P'}}(P_n)&=&3\sum\limits_{r=1}^{\lceil n/2\rceil}
\left(\frac{n^2+n(2r-3)-2r(r-1)}{n(n-1)(2n-1)}\right)\\
& &\cdot\log_2\left(\frac{2n(n-1)(2n-1)}
{3n^2+3n(2r-3)-6r(r-1)}\right).
\end{eqnarray*}
\end{thm}

In \cite{DS}, the authors presented explicit bounds or information
inequalities for any connected graph if the measure is based on the
information function using $j$-spheres, i.e., $f=f_P$ or
$f=f_{P'}$.
\begin{thm}\cite{DS}
Let $G=(V,E)$ be a connected graph on $n$ vertices. Then we infer
the following bounds:
\begin{eqnarray*}
I_{f_P}(G)&\leq&\left\{
\begin{array}{ll}
\beta^X\log_2(n\cdot\beta^X),\ &\text{if}\ \beta>1,\\
\beta^{-X}\log_2(n\cdot\beta^{-X}),\ &\text{if}\ \beta<1,
\end{array}
\right.\\
I_{f_P}(G)&\geq&\left\{
\begin{array}{ll}
\beta^X\log_2(n\cdot\beta^X),\ &\text{if}\ \left(\frac{1}{n}\right)^{\frac{1}{X}}\leq \beta\leq 1,\\
\beta^{-X}\log_2(n\cdot\beta^{-X}),\ &\text{if}\ 1\leq\beta\leq n^{\frac{1}{X}},\\
0,\ &\text{if}\ 0<\beta\leq\left(\frac{1}{n}\right) ^{\frac{1}{X}}\
\text{or}\ \beta\geq n^{\frac{1}{X}},
\end{array}
\right.
\end{eqnarray*}
where $X=(c_{max}-c_{min})(n-1)$ with $c_{max}=max\{c_j: 1\leq j\leq \rho(G)\}$ and $c_{min}=min\{c_j: 1\leq j\leq \rho(G)\}$.
\end{thm}

\begin{thm}\cite{DS}
Let $G=(V,E)$ be a connected graph on $n$ vertices. Then we infer the following bounds:
\begin{eqnarray*}
I_{f_{P'}}(G)&\leq&\frac{c_{max}}{c_{min}}\log_2
\left(\frac{n\cdot c_{max}}{c_{min}}\right),\\
I_{f_{P'}}(G)&\geq&\left\{
\begin{array}{ll}
0,\ &\text{if}\ n\leq\frac{c_{max}}{c_{min}},\\
\frac{c_{min}}{c_{max}}\log_2 \left(\frac{n\cdot
c_{min}}{c_{max}}\right), \ &\text{if}\ n>\frac{c_{max}}{c_{min}},
\end{array}
\right.
\end{eqnarray*}
where $c_{max}=max\{c_j: 1\leq j\leq \rho(G)\}$ and $c_{min}=min\{c_j: 1\leq j\leq \rho(G)\}$.
\end{thm}

Let $I_{f_1}(G)$ and $I_{f_2}(G)$ be entropies of graph $G$ defined
using the information functions $f_1$ and $f_2$, respectively.
Further, we define another function $f(v)=c_1f_1(v)+c_2f_2(v),\
v\in V$. In the following, we will give the relations between the
graph entropy $I_{f}(G)$ and the entropies $I_{f_1}(G)$ and
$I_{f_2}(G)$ which were found and proved by Dehmer and Sivakumar
\cite{DS}.
\begin{thm}\cite{DS}
Suppose $f_1(v)\leq f_2(v)$ for all $v\in V$. Then $I_{f}(G)$ can be
bounded by $I_{f_1}(G)$ and $I_{f_2}(G)$ as follows:
\begin{eqnarray*}
I_{f}(G)&\geq&
\frac{(c_1+c_2)A_1}{A}\left(I_{f_1}(G)-\log_2\frac{c_1A_1}{A}\right)
-\frac{c_2(c_1+c_2)A_2}{c_1A\ln(2)},\\
I_{f}(G)&\leq&
\frac{(c_1+c_2)A_2}{A}\left(I_{f_2}(G)-\log_2\frac{c_2A_2}{A}\right),
\end{eqnarray*}
where $A=c_1A_1+c_2A_2$, $A_1=\sum\limits_{v\in V}f_1(v)$ and $A_2=\sum\limits_{v\in V}f_2(v)$.
\end{thm}

\begin{thm}\cite{DS}
Given two information functions $f_1(v),f_2(v)$ such that
$f_1(v)\leq f_2(v)$ for all $v\in V$, then
\begin{equation*}
I_{f_1}(G)\leq \frac{A_2}{A_1}I_{f_2}(G)+
\log_2\frac{A_1}{A_1+A_2}-\frac{A_2}{A_1}\log_2\frac{A_2}{A_1+A_2}
+\frac{A_2\log_2e}{A_1}
\end{equation*}
where $A_1=\sum\limits_{v\in V}f_1(v)$ and $A_2=\sum\limits_{v\in
V}f_2(v)$.
\end{thm}

The next theorem gives another bound for $I_f(G)$ in terms of both
$I_{f_1}(G)$ and $I_{f_2}(G)$ by using the concavity property of the
logarithmic function.
\begin{thm}\cite{DS}
Let $f_1(v)$ and $f_2(v)$ be two arbitrary functions defined on a
graph $G$. If $f(v)=c_1f_1(v)+c_2f_2(v)$ for all $v\in V$, we infer
\begin{eqnarray*}
I_{f}(G)&\geq&\frac{c_1A_1}{A}\left[I_{f_1}(G)
-\log_2\frac{c_1A_1}{A}\right]+\frac{c_2A_2}{A}\left[I_{f_2}(G)
-\log_2\frac{c_2A_2}{A}\right]-\log_2e,\\
I_{f}(G)&\leq&\frac{c_1A_1}{A}\left[I_{f_1}(G)
-\log_2\frac{c_1A_1}{A}\right]+\frac{c_2A_2}{A}\left[I_{f_2}(G)
-\log_2\frac{c_2A_2}{A}\right],
\end{eqnarray*}
where $A=c_1A_1+c_2A_2$, $A_1=\sum\limits_{v\in V}f_1(v)$ and
$A_2=\sum\limits_{v\in V}f_2(v)$.
\end{thm}

The following theorem is a straightforward extension of the previous
statement. Here, an information function is expressed as a linear
combination of $k$ arbitrary information functions.
\begin{cor}\cite{DS}
Let $k\geq 2$ and $f_1(v),f_2(v),\cdots,f_k(v)$ be arbitrary
functions defined on a graph $G$. If
$f(v)=c_1f_1(v)+c_2f_2(v)+\cdots+c_kf_k(v)$ for all $v\in V$, we
infer
\begin{eqnarray*}
I_{f}(G)&\geq&\sum\limits_{i=1}^{k}\left\{\frac{c_iA_i}{A}
\left[I_{f_i}(G)-\log_2\frac{c_iA_i}{A}\right]\right\}-(k-1)\log_2e,\\
I_{f}(G)&\leq&\sum\limits_{i=1}^{k}\left\{\frac{c_iA_i}{A}
\left[I_{f_i}(G)-\log_2\frac{c_iA_i}{A}\right]\right\},
\end{eqnarray*}
where $A=\sum\limits_{i=1}^{k}c_iA_i$, $A_j=\sum\limits_{v\in
V}f_j(v)$ for $1\leq j\leq k$.
\end{cor}

Let $G_1=(V_1,E_1)$ and $G_2=(V_2,E_2)$ be two arbitrary connected
graphs on $n_1$ and $n_2$ vertices, respectively. The \emph{union of
the graphs} $G_1\cup G_2$ is the disjoint union of $G_1$ and $G_2$.
The \emph{join of the graphs} $G_1+G_2$ is defined as the graph
$G=(V,E)$ with vertex set $V=V_1\cup V_2$ and edge set $E=E_1\cup
E_2\cup \{(x,y):x\in V_1,y\in V_2\}$. In the following, we will
state the results of entropy $I_{f}(G)$ based on union of graphs and
join of graphs.
\begin{thm}\cite{DS}
Let $G=(V,E)=G_1\cup G_2$ be the disjoint union of graphs
$G_1=(V_1,E_1)$ and $G_2=(V_2,E_2)$. Let $f$ be an arbitrary
information function. Then
\begin{equation*}
I_{f}(G)=\frac{A_1}{A}\left(I_{f}(G_1)-\log_2\frac{A_1}{A}\right)
+\frac{A_2}{A}\left(I_{f}(G_2)-\log_2\frac{A_2}{A}\right)
\end{equation*}
where $A=A_1+A_2$ with $A_1=\sum\limits_{v\in V_1}f_{G_1}(v)$ and
$A_2=\sum\limits_{v\in V_2}f_{G_2}(v)$.
\end{thm}

As an immediate generalization of the previous theorem by taking $k$
disjoint graphs into account, we have the following corollary.
\begin{cor}\cite{DS} Let
$G_1=(V_1,E_1),G_2=(V_2,E_2),\cdots,G_k=(V_k,E_k)$ be $k$ arbitrary
connected graphs on $n_1,n_2,\cdots,n_k$ vertices, respectively. Let
$f$ be an arbitrary information function. Let $G=(V,E)=G_1\cup
G_2\cup\cdots\cup G_k$ be the disjoint union of graphs $G_i$. Then
\begin{equation*}
I_{f}(G)=\sum\limits_{i=1}^{k}\left\{\frac{A_i}{A
}\left(I_{f}(G_i)-\log_2\frac{A_i}{A}\right)\right\}
\end{equation*}
where $A=A_1+A_2+\cdots+A_k$ with $A_i=\sum\limits_{v\in
V_i}f_{G_i}(v)$ for $1\leq i\leq k$.
\end{cor}

Next we focus on the value of $I_{f_P}(G)$ and $I_{f_{P'}}(G)$
depending on the join of graphs.
\begin{thm}\cite{DS}
Let $G=(V,E)=G_1+G_2$ be the join of graphs $G_1=(V_1,E_1)$ and
$G_2=(V_2,E_2)$, where $|V_i|=n_i,\ i=1,2$. The graph entropy
$I_{f_P}(G)$ can then be expressed in terms of $I_{f_P}(G_1)$ and
$I_{f_P}(G_2)$ as follows:
\begin{equation*}
I_{f_P}(G)=\frac{A_1\beta^{c_1n_2}}{A}\left(
I_{f_P}(G_1)-\log_2\frac{A_1\beta^{c_1n_2}}{A}\right)
+\frac{A_2\beta^{c_1n_1}}{A}\left(
I_{f_P}(G_2)-\log_2\frac{A_2\beta^{c_1n_2}}{A}\right)
\end{equation*}
where $A=A_1\beta^{c_1n_2}+A_2\beta^{c_1n_2}$ with
$A_1=\sum\limits_{v\in V_1}f_{G_1}(v)$ and $A_2=\sum\limits_{v\in
V_2}f_{G_2}(v)$.
\end{thm}

\begin{thm}\cite{DS}
Let $G=(V,E)=G_1+G_2$ be the join of graphs $G_1=(V_1,E_1)$ and
$G_2=(V_2,E_2)$, where $|V_i|=n_i,\ i=1,2$. Then
\begin{equation*}
I_{f_{P'}}(G)\geq\frac{A_1}{A}\left(
I_{f_{P'}}(G_1)-\log_2\frac{A_1}{A}\right)
+\frac{A_2}{A}\left(I_{f_{P'}}(G_2)-\log_2\frac{A_2}{A}\right)
-\frac{2c_1n_1n_2}{A\ln(2)}
\end{equation*}
where $A=2c_1n_1n_2+A_1+A_2$ with $A_1=\sum\limits_{v\in
V_1}f_{G_1}(v)$ and $A_2=\sum\limits_{v\in V_2}f_{G_2}(v)$.
\end{thm}

Furthermore, an alternate set of bounds have been achieved in
\cite{DS}.
\begin{thm}\cite{DS}
Let $G=(V,E)=G_1+G_2$ be the join of graphs $G_1=(V_1,E_1)$ and
$G_2=(V_2,E_2)$, where $|V_i|=n_i,\ i=1,2$. Then
\begin{eqnarray*}
I_{f_{P'}}(G)&\leq&\frac{A_1}{A}\left(
I_{f_{P'}}(G_1)-\log_2\frac{A_1}{A}\right)
+\frac{A_2}{A}\left(I_{f_{P'}}(G_2)-\log_2\frac{A_2}{A}\right)
-\frac{c_1n_1n_2}{A}\log_2\frac{c_1^2n_1n_2}{A^2},\\
I_{f_{P'}}(G)&\geq&\frac{A_1}{A}\left(
I_{f_{P'}}(G_1)-\log_2\frac{A_1}{A}\right)
+\frac{A_2}{A}\left(I_{f_{P'}}(G_2)-\log_2\frac{A_2}{A}\right)\\
& &-\frac{c_1n_1n_2}{A}\log_2\frac{c_1^2n_1n_2}{A^2}-\log_2e,
\end{eqnarray*}
where $A=2c_1n_1n_2+A_1+A_2$ with $A_1=\sum\limits_{v\in
V_1}f_{G_1}(v)$ and $A_2=\sum\limits_{v\in V_2}f_{G_2}(v)$.
\end{thm}

\subsection{Extremal properties of degree-based and distance-based graph entropies}

Many graph invariants have been used to construct entropy-based
measures to characterize the structure of complex networks or deal
with inferring and characterizing relational structures of graphs in
discrete mathematics, computer science, information theory,
statistics, chemistry, biology, etc. In this section, we will state
the extremal properties of graph entropies that are based on
information functions $f^l_d(v_i)=d^l_i$ and
$f^n_k(v_i)=n_k(v_i)$, respectively, where $l$ is an arbitrary real
number and $n_k(v_i)$ is the number of vertices with distance $k$ to
$v_i$, $1\leq k\leq \rho(G)$.

In this section, we assume that $G=(V,E)$ is a simple connected
graph with $n$ vertices and $m$ edges. By applying Equation
\ref{eqCC} in Definition \ref{defGG}, we can obtain two special
graph entropies based on information functions $f^l_d$ and
$f^n_k$.
\begin{eqnarray*}
I_{f^l_d}(G)&:=&-\sum\limits_{i=1}^{n}
\frac{d^l_i}{\sum_{j=1}^{n}d^l_j}
\log\frac{d^l_i}{\sum_{j=1}^{n}d^l_j}
=\log\left(\sum_{i=1}^{n}d^l_i\right)
-\sum\limits_{i=1}^{n}\frac{d^l_i}{\sum_{j=1}^{n}d^l_j}
\log d^l_i,\\
I_{f^n_k}(G)&:=&-\sum\limits_{i=1}^{n}
\frac{n_k(v_i)}{\sum_{j=1}^{n}n_k(v_j)}
\log\left(\frac{n_k(v_i)}{\sum_{j=1}^{n}n_k(v_j)}\right)\\
&=&\log\left(\sum_{i=1}^{n}n_k(v_i)\right)
-\frac{1}{\sum_{j=1}^{n}n_k(v_j)}\sum\limits_{i=1}^{n} n_k(v_i)\log
n_k(v_i).
\end{eqnarray*}
The entropy $I_{f^l_d}(G)$ is based on an information function by
using degree powers, which is one of the most important graph
invariants and has been proved useful in information theory, social
networks, network reliability and mathematical chemistry
\cite{BN1,BN2}. In addition, the sum of degree powers has received
considerable attention in graph theory and extremal graph theory,
which is related to the famous Ramsey problem
\cite{Goodman1,Goodman2}. Meanwhile, the entropy $I_{f^n_k}(G)$
relates to a new information function, which is the number of
vertices with distance $k$ to a given vertex. Distance is one of the
most important graph invariants. For a given vertex $v$ in a graph,
the number of pairs of vertices with distance three, which is
related to the clustering coefficient of networks \cite{CRV}, is
also called the Wiener polarity index introduced by Wiener
\cite{Wiener}.

Since $\sum\limits_{i=1}^{n}d_i=2m$, we have
\begin{equation*}
I_{f^1_d}=\log(2m)-\frac{1}{2m}\sum\limits_{i=1}^{n}(d_i\log d_i).
\end{equation*}
In \cite{CDS}, the authors focused on extremal properties of graph
entropy $I_{f^1_d}(G)$ and obtained the maximum and minimum
entropies for certain families of graphs, i.e., trees, unicyclic
graphs, bicyclic graphs, chemical trees and chemical graphs.
Furthermore, they proposed some conjectures for extremal values of
those measures of trees.
\begin{thm}\cite{CDS}
Let $T$ be a tree on $n$ vertices. Then we have $I_{f^1_d}(T)\leq
I_{f^1_d}(P_n)$, the equality holds if and only if $T\cong P_n$;
$I_{f^1_d}(T)\geq I_{f^1_d}(S_n)$, the equality holds if and only if
$T\cong S_n$.
\end{thm}

A \emph{dendrimer} is a tree with $2$ additional parameters, the
progressive degree $p$ and the radius $r$. Every internal vertex of
the tree has degree $p+1$. In \cite{CDES}, the authors obtained the
following result.
\begin{thm}\cite{CDES}
Let $D$ be a dendrimer with $n$ vertices. The star graph and path
graph attain the minimal and maximal value of $I_{f^1_d}(D)$,
respectively.
\end{thm}

\begin{thm}\cite{CDS}
Let $G$ be a unicyclic graph with $n$ vertices. Then we have
$I_{f^1_d}(G)\leq I_{f^1_d}(C_n)$, the equality holds if and only if
$G\cong C_n$; $I_{f^1_d}(G)\geq I_{f^1_d}(S^+_n)$, the equality
holds if and only if $G\cong S^+_n$.
\end{thm}

Denote by $G^*$ and $G^{**}$ the bicyclic graphs with degree
sequence $[3^2,2^{n-2}]$ and $[n-1,3,2^2,1^{n-4}]$, respectively.
\begin{thm}\cite{CDS}
Let $G$ be a bicyclic graph of order $n$. Then we have
$I_{f^1_d}(G)\leq I_{f^1_d}(G^*)$, the equality holds if and only if
$G\cong G^*$; $I_{f^1_d}(G)\geq I_{f^1_d}(G^{**})$, the equality
holds if and only if $G\cong G^{**}$.
\end{thm}

In chemical graph theory, a chemical graph is a representation of
the structural formula of a chemical compound in terms of graph
theory. In this case, a graph corresponds to a chemical structural
formula, in which a vertex and an edge correspond to an atom and a
chemical bond, respectively. Since carbon atoms are $4$-valent, we
obtain graphs in which no vertex has degree greater than four. A
chemical tree is a tree $T$ with maximum degree at most four. We
call chemical graphs with $n$ vertices and $m$ edges
$(n,m)$-chemical graphs. For a more thorough introduction on
chemical graphs, we refer to \cite{BR1,Trinajstic}.

Let $T^*$ be a tree with $n$ vertices and $n-2=3a+i,\ i=0,1,2$,
whose degree sequence is $[4^a,i+1,1^{n-a-1}]$. Let $G_1$ be the
$(n,m)$-chemical graph with degree sequence $[d_1,d_2,\cdots,d_n]$
such that $|d_i-d_j|\leq 1$ for any $i\neq j$ and $G_2$ be an
$(n,m)$-chemical graph with at most one vertex of degree $2$ or $3$.
\begin{thm}\cite{CDS}
Let $T$ be a chemical tree of order $n$ such that $n-2=3a+i,\
i=0,1,2$. Then we have $I_{f^1_d}(T)\leq I_{f^1_d}(P_n)$, the
equality holds if and only if $T\cong P_n$; $I_{f^1_d}(T)\geq
I_{f^1_d}(T^{*})$, the equality holds if and only if $T\cong T^{*}$.
\end{thm}

\begin{thm}\cite{CDS}
Let $G$ be an $(n,m)$-chemical graph. Then we have $I_{f^1_d}(G)\leq
I_{f^1_d}(G_1)$, the equality holds if and only if $G\cong G_1$;
$I_{f^1_d}(G)\geq I_{f^1_d}(G_2)$, the equality holds if and only if
$G\cong G_2$.
\end{thm}

By performing numerical experiments, the authors \cite{CDS} proposed
the following conjecture while several attempts to prove the
statement by using different methods failed.
\begin{con}\cite{CDS}
Let $T$ be a tree with $n$ vertices and $l>0$. Then we have
$I_{f^l_d}(T)\leq I_{f^l_d}(P_n)$, the equality holds if and only if
$T\cong P_n$; $I_{f^l_d}(T)\geq I_{f^l_d}(S_n)$, the equality holds
if and only if $T\cong S_n$.
\end{con}

Furthermore, Cao and Dehmer \cite{CD} extended the results performed
in \cite{CDS}. The authors explored the extremal values of
$I_{f^l_d}(G)$ and the relations between this entropy and the sum of
degree powers for different values of $l$. In addition, they
demonstrated those results by generating numerical results using
trees with $11$ vertices and connected graphs with $7$ vertices,
respectively.
\begin{thm}\cite{CD}
Let $G$ be a graph with $n$ vertices. Denote by $\delta$ and
$\Delta$ the minimum degree and maximum degree of $G$, respectively.
Then we have
\begin{equation*}
\log\left(\sum\limits_{i=1}^{n}d_i^l\right)-l\log\Delta\leq I_{f^l_d}(G)
\leq \log\left(\sum\limits_{i=1}^{n}d_i^l\right)-l\log\delta.
\end{equation*}
\end{thm}

The following corollary can be obtained directly from the above theorem.
\begin{cor}\cite{CD}
If $G$ is a $d$-regular graph, then $I_{f^l_d}(G)=\log n$ for any $l$.
\end{cor}

Observe that if $G$ is regular, then $I_{f^l_d}(G)$ is a function
only on $n$. For the trees with $11$ vertices and connected graphs
with $7$ vertices, the authors \cite{CD} gave numerical results on
$\sum_{i=1}^{n}d_i^l$ and $I_{f^l_d}(G)$, which gives support for
the following conjecture.
\begin{con}\cite{CD}
For $l>0$, $I_{f^l_d}(G)$ is a monotonously increasing function on
$l$ for connected graphs.
\end{con}

In \cite{ChDS}, the authors discuss the extremal properties of the
graph entropy $I_{f^n_k}(G)$ thereof leading to a better
understanding of this new information-theoretic quantity. For $k=1$,
$I_{f^n_1}(G)=\log(2m)-\frac{1}{2m}\cdot \sum_{i=1}^{n}d_i\log d_i$
because $n_1(v_i)=d_i$ and $\sum_{i=1}^{n}d_i=2m$. Denote by
$p_k(G)$ the number of geodesic paths with length $k$ in graph $G$.
Then we have $\sum_{i=1}^{n}n_k(v_i)=2p_k(G)$, since each path of
length $k$ is counted twice in $\sum_{i=1}^{n}n_k(v_i)$. Therefore,
\begin{equation*}
I_{f^n_k}(G)=\log(2p_k(G))
-\frac{1}{2p_k(G)}\cdot\sum\limits_{i=1}^{n}
n_k(v_i)\log n_k(v_i).
\end{equation*}
As is known to all, there are some good algorithms for finding
shortest paths in a graph. From this aspect, the authors obtained
the following result first.
\begin{pro}\cite{ChDS}
Let $G$ be a graph with $n$ vertices. For a given integer $k$, the
value of $I_{f^n_k}(G)$ can be computed in polynomial time.
\end{pro}

Let $T$ be a tree with $n$ vertices and
$V(T)=\{v_1,v_2,\cdots,v_n\}$. In the following, we present the
properties of $I_{f^n_k}(T)$ for $k=2$ proved by Chen, Dehmer and
Shi \cite{ChDS}. By some elementary calculations, the authors
\cite{ChDS} found that
\begin{eqnarray*}
I_{f^n_2}(T)&=&\log\left(\sum_{i=1}^{n}d_i^2-2(n-1)\right)
-\frac{\sum_{i=1}^{n}n_2(v_i)\log n_2(v_i)}
{\sum_{i=1}^{n}d_i^2-2(n-1)},\\
I_{f^n_2}(S_n)&=&\log(n-1),\\
I_{f^n_2}(P_n)&=&\log(n-2)+\frac{2}{n-2},\\
I_{f^n_2}(S_{\lfloor\frac{n}{2}\rfloor,\lceil\frac{n}{2}\rceil})&=&
\left\{\begin{array}{ll}
\log(n)\ &if\ n=2k,\\
\frac{3k-1}{2k}\log(k)-\frac{k-1}{2k}\log(k-1)+1\ &if\ n=2k+1,
\end{array}\right.\\
I_{f^n_2}(CS(n,t))&=&\log(t^2-3t+2n-2)\\
& &-\frac{2(n-t-3)+n_1\log t+(t-1)^2\log(t-1)}{t^2-3t+2n-2},
\ n-t\geq 3.
\end{eqnarray*}
Then they obtained the following result.
\begin{thm}\cite{ChDS}
Let $S_n,P_n,S_{\lfloor\frac{n}{2}\rfloor, \lceil\frac{n}{2}\rceil}$
be the star, the path and the balanced double star with $n$ vertices,
respectively. Then
\begin{equation*}
I_{f^n_2}(S_n)<I_{f^n_2}(P_n)<I_{f^n_2}(S_{\lfloor\frac{n}{2}\rfloor,
\lceil\frac{n}{2}\rceil}).
\end{equation*}
\end{thm}

Depending on the above extremal trees of $I_{f^n_2}(T)$, Chen,
Dehmer and Shi \cite{ChDS} proposed the following conjecture.
\begin{con}\cite{ChDS}
For a tree $T$ with $n$ vertices, the balanced double star and the
comet $CS(n,t_0)$ can attain the maximum and the minimum values of
$I_{f^n_2}(T)$, respectively.
\end{con}

\begin{figure}[h,t,b,p]
\begin{center}
\includegraphics[scale = 0.9]{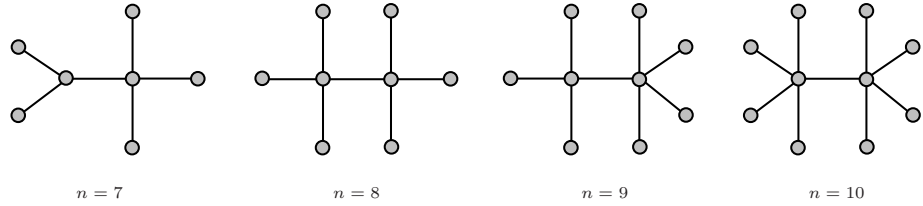}
\caption{The trees with maximum value of $I_{f^n_2}(T)$ among all trees with $n$ vertices for $7\leq n\leq 10$.} \label{fig1}
\end{center}
\end{figure}

\begin{figure}[h,t,b,p]
\begin{center}
\includegraphics[scale = 0.9]{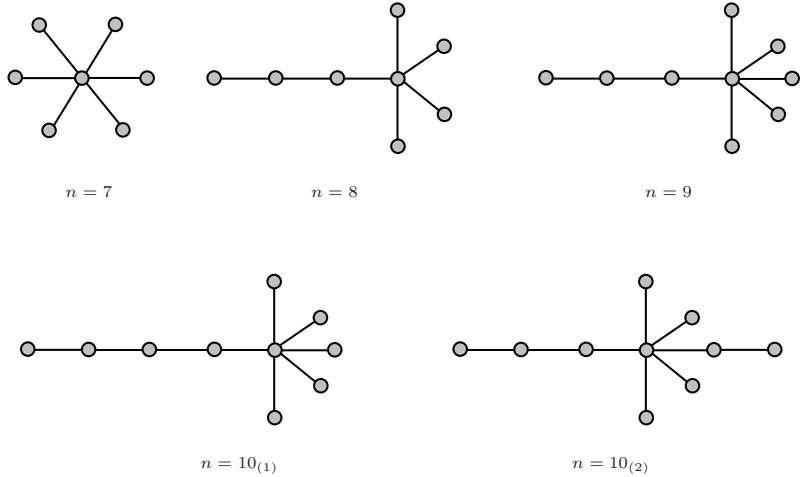}
\caption{The trees with minimum value of $I_{f^n_2}(T)$ among all trees with $n$ vertices for $7\leq n\leq 10$.} \label{fig2}
\end{center}
\end{figure}

By calculating the values $I_{f^n_2}(T)$ for $n=7,8,9,10$, the
authors obtained the trees with extremal values of entropy which are
shown in Figures \ref{fig1} and \ref{fig2}, respectively.

Observe that the extremal graphs for $n=10$ is not unique. From this
observation, they \cite{ChDS} obtained the following result.
\begin{thm}\cite{ChDS}
Let $CS(n,t)$ be a comet with $n-t\geq 4$. Denote by $T$ a tree
obtained from $CS(n,t)$ by deleting the leaf that is not adjacent to
the vertex of maximum degree and attaching a new vertex to one leaf
that is adjacent to the vertex of maximum degree. Then
$I_{f^n_2}(T)=I_{f^n_2}(CS(n,t))$.
\end{thm}

\subsection{Extremality of $I_{f^\lambda}(G)$, $I_{f^2}(G)$ $I_{f^3}(G)$ and entropy bounds for dendrimers}

In the setting of information-theoretic graph measures, we will
often consider a tuple $(\lambda_1,\lambda _2,\cdots,\lambda_k)$ of
nonnegative integers $\lambda_i\in N$. Let $G=(V,E)$ be a connected
graph with $|V|=n$ vertices. Here, we define
$f^\lambda(v_i)=\lambda_i$, for all $v_i\in V$. Next we define
$f^2,f^3$ as follows.
\begin{definition}\cite{DK}
Let $G=(V,E)$. For a vertex $v_i\in V$, we define
\begin{eqnarray*}
f^2(v_i)&:=&c_1|S_1(v_i,G)|+c_2|S_2(v_i,G)|+\cdots
+c_\rho|S_\rho(v_i,G)|,\\
& &c_k>0,\ 1\leq k\leq \rho,\ \alpha>0,\\
f^3(v_i)&:=&c_i\sigma(v_i),\quad c_k>0,\ 1\leq k\leq n,
\end{eqnarray*}
where $\sigma(v)$ and $S_j(v,G)$ are the eccentricity and the
$j$-sphere of vertex $v$, respectively.
\end{definition}

For the information function $f^2$, by applying the Equations
\ref{eqCC} and \ref{eqDD} in Definition \ref{defGG}, we can obtain
the following two entropy measures $I_{f^2}(G)$ and
$I_{f^2}^\lambda(G)$.
\begin{eqnarray*}
I_{f^2}(G)&:=&-\sum\limits_{i=1}^{|V|}\frac{f^2(v_i)}
{\sum_{j=1}^{|V|}f^2(v_j)}
\log \frac{f^2(v_i)}{\sum_{j=1}^{|V|}f^2(v_j)},\\
I_{f^2}^\lambda(G)&:=&\lambda\left(\log(|V|)
+\sum\limits_{i=1}^{|V|}\frac{f^2(v_i)}{\sum_{j=1}^{|V|}f^2(v_j)}
\log \frac{f^2(v_i)}{\sum_{j=1}^{|V|}f^2(v_j)}\right).
\end{eqnarray*}
In \cite{DBVG}, the authors proved that if the graph $G=(V,E)$ is
$k$-regular, then $I_{f^2}(G)=\log(|V|)$ and, hence,
$I_{f^2}^\lambda(G)=0$.

For our purpose, we will mainly use decreasing sequences $c_1,\cdots,c_{\rho(G)}$ of\\[2mm]
(1) \ constant decrease: $c_1:=S,c_2:=S-k,\cdots,c_{\rho(G)}:=S-(\rho(G)-1)k$,\\[2mm]
(2) \ quadratic decrease: $c_1:=S^2,c_2:=(S-k)^2,\cdots,c_{\rho(G)}:=(S-(\rho(G)-1)k)^2$,\\[2mm]
(3) \ exponential decrease: $c_1:=S,c_2:=Se^{-k},\cdots,c_{\rho(G)}:=Se^{-(\rho(G)-1)k}$.\\[2mm]
Intuitive choices for the parameters are $S=\rho(G)$ and $k=1$.

Applying the Equation \ref{eqCC} in Definition \ref{defGG}, we can
obtain three graph entropies as follows.
\begin{eqnarray*}
I(\lambda_1,\lambda_2,\cdots,\lambda_n)=I_{f^\lambda}(G)
&=&-\sum\limits_{i=1}^{n}\frac{\lambda_i}{\sum_{j=1}^{n}\lambda_j}
\log\frac{\lambda_i}{\sum_{j=1}^{n}\lambda_j},\\
I_{f^2}(G)&=&-\sum\limits_{i=1}^{n}\frac{f^2(v_i)}
{\sum_{j=1}^{n}f^2(v_j)}
\log\frac{f^2(v_i)}{\sum_{j=1}^{n}f^2(v_j)},\\
I_{f^3}(G)&=&-\sum\limits_{i=1}^{n}\frac{f^3(v_i)}
{\sum_{j=1}^{n}f^3(v_j)}
\log\frac{f^3(v_i)}{\sum_{j=1}^{n}f^3(v_j)}.
\end{eqnarray*}
As described in Definition \ref{defHH}, $p_i:=p^{f^\lambda}(v_i)=
\frac{\lambda_i}{\sum_{j=1}^{n}\lambda_j},\ i=1,2,\cdots,n$. We call
$p=(p_1,p_2,\cdots,p_n)$ be the \emph{probability distribution
vector}. Depending on probability distribution vector, we denote the entropy $I_{f^\lambda}(G)$ as $I_p(G)$ as well. Now we present some extremal properties of the entropy measure $I_{f^\lambda}(G)$, i.e.,
$I(\lambda_1,\lambda_2,\cdots,\lambda_n)$.
\begin{lem}\cite{DK}
If

(i) \ $\lambda_m+x\leq\frac{\Sigma_m}{n-1}$ and $x\geq 0$, or

(ii) \
$(\lambda_m+x)\Sigma_m\geq\sum_{i\in\{1,\cdots,n\}-m}\lambda_i^2$
and $-\lambda_m<x<0$,

then
\begin{equation*}
I(\lambda_1,\cdots,\lambda_{m-1},
\lambda_m,\lambda_{m+1},\cdots,\lambda_n)\leq
I(\lambda_1,\cdots,\lambda_{m-1},
\lambda_m+x,\lambda_{m+1},\cdots,\lambda_n);
\end{equation*}

on the other hand, if

(iii) \ $\lambda_m\leq\frac{\Sigma_m}{n-1}$ and $-\lambda_m<x<0$, or

(iv) \ $\lambda_m\Sigma_m\geq\sum_{i\in\{1,\cdots,n\}-m}\lambda_i^2$
and $x>0$,

then
\begin{equation*}
I(\lambda_1,\cdots,\lambda_{m-1},
\lambda_m,\lambda_{m+1},\cdots,\lambda_n)\geq
I(\lambda_1,\cdots,\lambda_{m-1},
\lambda_m+x,\lambda_{m+1},\cdots,\lambda_n),
\end{equation*}
where $x\geq -\lambda_m$, $\Sigma=\sum_{j=1}^{n}\lambda_j$ and $\Sigma_m=\sum_{j\in\{1,\cdots,n\}-m}\lambda_j$.
\end{lem}

Let $p=(p_1,p_2,\cdots,p_n)$ be the original probability
distribution vector and
$\bar{p}=(\bar{p}_1,\bar{p}_2,\cdots,\bar{p}_n)$ be the changed one,
both ordered in increasing order. Further, let $\Delta
p=\bar{p}-p=(\delta_1,\cdots,\delta_n)$ where
$\delta_1,\cdots,\delta_n\in \mathbb{R}$. Obviously,
$\sum_{i=1}^{n}\delta_i=0$.
\begin{lem}\label{lemAA}\cite{DK}
If

(i) \ there exists a $k$ such that for all $1\leq i\leq k$,
$\delta_i\leq 0$ and for all $k+1\leq i\leq n$, $\delta_i\geq 0$ or,
more generally, if

(ii) \ $\sum_{i=1}^{\ell}\delta_i\leq 0$ for all $\ell=1,\cdots,n$,

then
\begin{equation*}
I_{\bar{p}}(G)\leq I_p(G).
\end{equation*}
\end{lem}

\begin{lem}\cite{DK}
For two probability distribution vectors $\bar{p}$ and $p$
fulfilling condition (ii) of Lemma \ref{lemAA}, we have
\begin{equation*}
I_p(G)-I_{\bar{p}}(G)\geq\sum\limits_{i=1}^{n}\delta_i\log p_i
\end{equation*}
where $\delta_i$ are the entries of $\Delta p=\bar{p}-p$.
\end{lem}

\begin{lem}\cite{DK}
Assume that for two probability distribution vectors $p$ and
$\bar{p}$, the opposite of condition (ii) in Lemma \ref{lemAA} is
true, that is $\sum_{i=\ell}^{n}\delta_i\geq 0$ for all
$\ell=1,\cdots,n$. Then
\begin{equation*}
0>I_p(G)-I_{\bar{p}}(G)\geq\sum\limits_{i:\delta_i<0}
\delta_i\log(p_i-\rho)+\sum\limits_{i:\delta_i>0}\delta_i\log(p_i+\rho)
\end{equation*}
where $\rho=max_{i\in \{2,\cdots,n\}}(p_i-p_{i-1})$.
\end{lem}

\begin{pro}\cite{DK}
For two probability distribution vectors $p$ and $\bar{p}$ with
$\sum_{i=1}^{\ell}\delta_i\leq 0$ for all $\ell$ in
$\{0,\cdots,\ell_1-1\}\cup \{\ell_2,\cdots,n\} \ (1\leq
\ell_1<\ell_2\leq n)$, we have that
\begin{eqnarray*}
I_p(G)-I_{\bar{p}}(G)&\geq&\sum\limits_{i=1}^{\ell_1-1} \delta_i\log
p_i+\sum\limits_{i=\ell_2}^{n}\delta_i\log p_i
+\sum\limits_{i=\ell_1}^{\ell_2-1}\delta_i\log(p_i+\rho)\\
& &+\sum\limits_{i=1}^{\ell_1-1}
\delta_i\log\left(\frac{p_{\ell_1}-\rho}{p_{\ell_1}}\right)
+\sum\limits_{i=1}^{\ell_2-1}
\delta_i\log\left(\frac{p_{\ell_2}}{p_{\ell_2}+\rho}\right)
\end{eqnarray*}
where $\rho=max_{i\in \{2,\cdots,n\}}(p_i-p_{i-1})$. Hence, if
\begin{eqnarray*}
\sum\limits_{i=1}^{\ell_1-1}\delta_i\log p_i
+\sum\limits_{i=\ell_2}^{n}\delta_i\log p_i&\geq& -\left(
\sum\limits_{i=\ell_1}^{\ell_2-1}\delta_i\log(p_i+\rho)
+\sum\limits_{i=1}^{\ell_1-1}
\delta_i\log\left(\frac{p_{\ell_1}-\rho}{p_{\ell_1}}\right)\right.\\
& &\left.+\sum\limits_{i=1}^{\ell_2-1}
\delta_i\log\left(\frac{p_{\ell_2}}{p_{\ell_2}+\rho}\right)\right),
\end{eqnarray*}
it follows that
\begin{equation*}
I_p(G)\geq I_{\bar{p}}(G).
\end{equation*}
\end{pro}

In the following, we will show some results \cite{CDES,DK} regarding
the maximum and minimum entropy by using certain families of graphs.

As in every tree, a dendrimer has one (monocentric dendrimer) or two
(dicentric dendrimer) central vertices, the radius $r$ denotes the
(largest) distance from an external vertex to the (closer) center.
If all external vertices are at distance $r$ from the center, the
dendrimer is called \emph{homogeneous}. Internal vertices different
from the central vertices are called \emph{branching nodes} and are
said to be \emph{on the $i$-th orbit} if their distance to the
(nearer) center is $r$.

Let $D_n$ denote a homogeneous dendrimer on $n$ vertices with radius
$r$ and progressive degree $p$, and let $z$ be its (unique) center.
Further denote by $V_i(D_n)$ the set of vertices in the $i$-th
orbit. Now we consider the function $f^3(v_i)=c_i\sigma(v_i)$,
where $c_i=c_j$ for $v_i,v_j\in V_i$. We denote $\bar{c}_i=c(v),\
v\in V_i$.
\begin{lem}\cite{DK}
For $\bar{c}_i=1$ with $i=0,\cdots,n$, the entropy fulfills
\begin{equation*}
\log n-\frac{1}{4\ln 2}\leq I_{f^3}(D_n)\leq\log n.
\end{equation*}
For $\bar{c}_i=r-i+1$ with $i=0,\cdots,n$, we have
\begin{equation*}
\log n-\frac{(r-1)^2}{4\ln 2(r+1)}\leq I_{f^3}(D_n)=\log n.
\end{equation*}
In general, for weight sequence $\bar{c}(i)$, $i=0,\cdots,r$, where
$c(i)(r+i)$ is monotonic in $i$, we have
\begin{equation*}
\log n-\frac{(\rho-1)^2}{2\rho\ln 2}\leq I_{f^3}(D_n)\leq\log n
\end{equation*}
where $\rho=\frac{c(1)}{2c(r)}$ for decreasing and
$\rho=\frac{2c(r)}{c(1)}$ for increasing sequences. The latter
estimate is also true for any sequence $c(i)$, when
$\rho=\frac{max_i(c(i)(r+i))}{min_j(c(j)(r+j))}$.
\end{lem}

\begin{lem}\cite{DK}
For dendrimers, the entropy $I_{f^3}(D_n)$ is of order $\log n$ as $n$ tends to infinity.
\end{lem}

By performing numerical experiments, Dehmer and Kraus \cite{DK}
raised the following conjecture and also gave some ideas on how to
prove it.
\begin{con}\cite{DK}
Let $D$ be a dendrimer on $n$ vertices. For all sequences
$\bar{c}_0\geq \bar{c}_1\geq \cdots\geq \bar{c}_r$, the star graph
($r=1,p=n-2$) have maximal and the path graph ($r=\lceil n-1/2\rceil,p=1$)
have minimal values of entropy $I_{f^3}(D)$.
\end{con}

Additionally, in \cite{CDES}, the authors proposed another conjecture which is stated as follows.
\begin{con}\cite{CDES}
Let $D$ be a dendrimer on $n$ vertices. For all sequences $c_i=c_j$
with $i\neq j$, the star graph ($r=1,p=n-2$) has the minimal value
of entropy $I_{f^3}(D)$.
\end{con}

Let $G$ be a generalized tree with hight $h$ which is defined as in
Section \ref{subsAA}. Denote by $|V|$ and $|V_i|$ the total number
of vertices and the number of vertices on the $i$-th level,
respectively. A probability distribution based on the vertices of
$G$ is assigned as follows:
\begin{equation*}
p_{i}^{V'}=\frac{|V_i|}{|V|-1}.
\end{equation*}
Then another entropy of a generalized tree $G$ is defined by
\begin{equation*}
I^{V'}(G)=-\sum\limits_{i=1}^{h}p_{i}^{V'}\log(p_{i}^{V'}).
\end{equation*}
Similarly, denote by $|E|$ and $|E_i|$ the total number of edges and
the number of edges on the $i$-th level, respectively. A probability
distribution based on the edges of $G$ is assigned as follows:
\begin{equation*}
p_{i}^{E'}=\frac{|E_i|}{|E|-1}.
\end{equation*}
Then another entropy of a generalized tree $G$ is defined by
\begin{equation*}
I^{E'}(G)=-\sum\limits_{i=1}^{h}p_{i}^{E'}\log(p_{i}^{E'}).
\end{equation*}

Now we give some extremal properties \cite{CDES} of $I^{V'}(D)$ and $I^{E'}(D)$, where $D$ is a dendrimer.
\begin{thm}\cite{CDES}
Let $D$ be a dendrimer on $n$ vertices. The star graph attains the
minimal value of $I^{V'}(D)$ and $I^{E'}(D)$, and the dendrimer with
parameter $t=t_0$ attains the maximal value of $I^{V'}(D)$ and
$I^{E'}(D)$, where $t=t_0\in(1,n-2)$ is the integer which is closest
to the root of the equation
\begin{equation*}
\frac{n}{n-1}\ln\left(\frac{nt-n+2}{t+1}\right)
-\ln\left(\frac{t(t+1)}{n-1}\right)-\frac{2t}{t+1}=0.
\end{equation*}
\end{thm}

According to Rashevsky \cite{Rashevsky}, $|X_i|$ denotes the number
of topologically equivalent vertices in the $i$-th vertex orbit of
$G$, where $k$ is the number of different orbits. Suppose
$|X|=|V|-1$. Then the probability of $X_i$ can be expressed as
$p_{i}^{V'}=\frac{|V_i|}{|V|-1}$. Therefore, by applying Equations
\ref{eqKK}, \ref{eqII}, \ref{eqJJ} in Definition \ref{defII}, we can
obtain the entropies as follows:
\begin{eqnarray*}
&(i)&\ I^5(G):=\sum\limits_{i=1}^{k}p_i^{V'}(1-p_i^{V'}),\\
&(ii)&\ I^1_\alpha(G):=\frac{1}{1-\alpha}\log\left(
\sum\limits_{i=1}^{k}\left(p_i^{V'}\right)^\alpha\right),\ \alpha\neq 1,\\
&(iii)&\ I^3_\alpha(G):=\frac{\sum\limits_{i=1}^{k}\left(
p_i^{V'}\right)^\alpha-1}{2^{1-\alpha}-1},\ \alpha\neq 1.
\end{eqnarray*}

\begin{thm}\cite{CDES}
Let $D$ be a dendrimer on $n$ vertices.

(i) \ The star graph and path graph attain the minimal and maximal
value of $I^5(D)$, respectively.

(ii) \ For $\alpha\neq 1$, the star graph and path graph attain the
minimal and maximal value of $I^1_\alpha(D)$, respectively.

(iii) \ For $\alpha\neq 1$, the star graph and path graph attain the
minimal and maximal value of $I^3_\alpha(D)$, respectively.
\end{thm}

Next we describe the algorithm for uniquely decomposing a graph
$G\in \mathcal{G}_{UC}$ into a set of undirected generalized trees
\cite{Dehmer2}.

\noindent \emph{\bf Algorithm 3.1}: A graph $G\in \mathcal{G}_{UC}$
with $|V|$ vertices can be locally decomposed into a set of
generalized trees as follows: Assign vertex labels to all vertices
from $1$ to $|V|$. These labels form the label set
$L_S=\{1,\cdots,|V|\}$. Choose a desired height of the trees that is
denoted by $h$. Choose an arbitrary label from $L_S$, e.g., $i$. The
vertex with this label is the root vertex of a tree. Now, perform
the following steps:

1. Calculate the shortest distance from the vertex $i$ to all other
vertices in the graph $G$, e.g., by the algorithm of Dijkstra; see
Dijkstra (1959).

2. The vertices with distance $k$ from the vertex $i$ are the vertices on the $k$-th level of the resulting generalized trees. Select all vertices of the graph up to distance $h$, including the connections between the vertices. Connections to vertices with distance $>h$ are deleted.

3. Delete the label $i$ from the label set $L_S$.

4. Repeat this procedure if $L_S$ is not empty by choosing an
arbitrary label from $L_S$; otherwise terminate.

Now we replace $p_{i}^{E'}=\frac{|E_i|}{|E|-1}$ by
$p_{i}^{E'}=\frac{|E_i|}{2|E|-d(r)}$, where $r$ is the root of the
generalized tree and $d(r)$ is the degree of $r$. Then we can obtain
a new $I^{E'}(G)$ which is defined similarly as above. Additionally,
we give another definition of the structural information content of
a graph as follows.
\begin{definition}\cite{Dehmer2}
Let $G\in \mathcal{G}_{UC}$ and $S_G^H:=\{H_1,H_2,\cdots,H_{|V|}\}$ be the associated set of generalized trees obtained from Algorithm 3.1. We now define the structural information content of $G$ by
\begin{equation*}
I^{V'}(G):=-\sum\limits_{i=1}^{|V|}I^{V'}(H_i)
\end{equation*}
and
\begin{equation*}
I^{E'}(G):=-\sum\limits_{i=1}^{|V|}I^{E'}(H_i).
\end{equation*}
\end{definition}

In \cite{Dehmer2}, Dehmer analyzed the time complexities for
calculating the entropies $I^{V'}(G)$ and $I^{E'}(G)$ depending on
the decomposition given by Algorithm 3.1.
\begin{thm}\cite{Dehmer2}
The overall time complexity to calculate $I^{V'}(G)$ and $I^{E'}(G)$
is finally $O(|V|^3+\sum_{i=1}^{|V|}|V_{H_i}|^2)$.
\end{thm}

Let $T_{n,d}$ be the family of trees of order $n$ with a fixed
diameter $d$. We call a tree consisting of a star on $n-d+1$
vertices together with a path of length $d-1$ attached to the
central vertex, \emph{a comet of order $n$ with tail length $d-1$},
and denote it by $C_{n,d-1}$. Analogously, we call a tree consisting
of a star on $n-d$ vertices together with $2$ paths of lengths
$\lfloor\frac{d}{2}\rfloor$ and $\lceil\frac{d}{2}\rceil$,
respectively, attached to the central vertex, \emph{a two-tailed
comet of order $n$} and denote it by
$C_{n,\lfloor\frac{d}{2}\rfloor,\lceil\frac{d}{2}\rceil}$.
\begin{thm}\cite{DK}
For every linearly or exponentially decreasing sequence
$c_1>c_2>\cdots>c_d$ with $d\geq 4$ as well as every quadratically
decreasing sequence with $d\geq 5$, for large enough $n$, the
probability distribution $q(n,d)$ of the $2$-tailed comet
$C_{n,\lfloor\frac{d}{2}\rfloor,\lceil\frac{d}{2}\rceil}$ is
majorities by the probability distribution $p(n,d)$ of the comet
$C_{n,d-1}$. This is equivalent to the fact that $\Delta p=q-p$
fulfills condition (ii) of Lemma \ref{lemAA}. Hence,
\begin{equation*}
I_{f^2}(C_{n,\lfloor\frac{d}{2}\rfloor,\lceil\frac{d}{2}\rceil})
\geq I_{f^2}(C_{n,d-1}).
\end{equation*}
\end{thm}

\begin{con}\cite{DK}
Among all trees $T_{n,d}$, with $d<<n$, the $2$-tailed comet
$C_{n,\lfloor\frac{n}{2}\rfloor,\lceil\frac{n}{2}\rceil}$ achieves
maximal value of the entropies $I_{f^2}(G)$ and $I_{f^3}(G)$.
\end{con}

\subsection{Sphere-regular graphs and the extremality entropies $I_{f^2}(G)$ and $I_{f^\sigma}(G)$}

Let $G=(V,E)$ be a connected graph with $|V|=n$ vertices. As we have
defined before, the information function
$f^2(v_i)=c_1|S_1(v_i,G)|+c_2|S_2(v_i,G)|+\cdots
+c_\rho|S_\rho(v_i,G)|$, where $c_k>0$, $1\leq k\leq \rho,\
\alpha>0$, and $S_j(v,G)$ is the $j$-sphere of the vertex $v$. Now
we define another information function.
\begin{definition}
The eccentricity-function $f^\sigma$ if defined by
\begin{equation*}
f^\sigma:\ V\rightarrow \mathbb{Z}:\quad f^\sigma(v)=\sigma(v).
\end{equation*}
\end{definition}

Applying the Equation \ref{eqCC} in Definition \ref{defGG}, we can
obtain the following two graph entropy measures \cite{KDS}.
\begin{equation*}
I_{f^2}(G):=-\sum\limits_{i=1}^{n}\frac{f^2(v_i)}{\sum_{j=1}^{n}
f^2(v_j)}\log \frac{f^2(v_i)}{\sum_{j=1}^{n}f^2(v_j)},
\end{equation*}
\begin{equation*}
I_{f^\sigma}(G):=-\sum\limits_{i=1}^{n}\frac{f^\sigma(v_i)}
{\sum_{j=1}^{n}f^\sigma(v_j)}\log \frac{f^\sigma(v_i)}{\sum_{j=1}^{n}f^\sigma(v_j)}.
\end{equation*}

In \cite{KDS}, the authors proposed the concept of sphere-regular.
\begin{definition}\cite{KDS}
We call a graph sphere-regular if there exist positive integers $s_1,\cdots,s_{\rho(G)}$, such that
\begin{equation*}
(|S_1(v,G)|,|S_2(v,G)|,\cdots,|S_{\rho(G)}(v,G)|)
=(s_1,\cdots,s_{\rho(G)})
\end{equation*}
for all vertices $v\in V$.
\end{definition}

In \cite{KDS}, the authors also tried to classify those graphs which
return maximal value of entropy $I_{f^2}(G)$ for the
sphere-function and an arbitrary decreasing weight sequence. In
the following, we state their results.
\begin{pro}\cite{KDS}
Every sphere-regular graph with $n$ vertices has maximum entropy $I_{f^2}(G)=\log n$.
\end{pro}

\begin{lem}\cite{KDS}
Sphere-regular graphs are the only maximal graphs for $I_{f^2}(G)$
when using a weight sequence such that there exist no numbers $a_i,\
i=1,\cdots,\rho(G)$, $a_i\in \mathbb{Z}$ with
$\sum_{i=1}^{\rho(G)}a_i=0$, where
\begin{equation*}
\sum_{j=1}^{\rho(G)}a_jc_j=0.
\end{equation*}
\end{lem}

\begin{thm}\cite{KDS}
There are maximal graphs with respect to $I_{f^2}(G)$ which are not sphere-regular.
\end{thm}

Next, we will present some restrictions on maximal graphs for
$I_{f^2}(G)$, which are valid for any decreasing weight sequence.
\begin{lem}\cite{KDS}
A graph of diameter $2$ is maximal for $I_{f^2}(G)$ if and only if it is sphere-regular.
\end{lem}

\begin{lem}\cite{KDS}
Maximal graphs for $I_{f^2}(G)$ cannot have unary vertices (vertices with degree $1$). Hence, in particular, trees cannot be maximal for $I_{f^2}(G)$.
\end{lem}

\begin{cor}\cite{KDS}
The last nonzero entries of the sphere-sequence of a vertex in a
maximal graph cannot be $2$ or more consecutive ones.
\end{cor}

\begin{lem}\cite{KDS}
A maximal graph for $I_{f^2}(G)$ different from the complete graph $K_n$ cannot contain a vertex of degree $n-1$.
\end{lem}

\begin{figure}[h,t,b,p]
\begin{center}
\includegraphics[scale = 0.9]{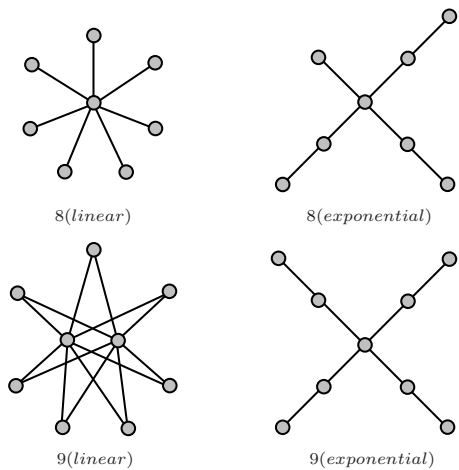}
\caption{Minimal graphs for $I_{f^2}(G)$ of orders $8$ and $9$.}
\label{fig3}
\end{center}
\end{figure}

In \cite{KDS}, the authors gave the minimal graphs for $I_{f^2}(G)$
of orders $8$ and $9$ by computations, which are depicted in Figure
\ref{fig3}. Its left graphs are minimal for the linear sequence and
the right ones are minimal for the exponential sequence.
Unfortunately, there is very little known about minimal entropy
graphs. And the authors gave the following conjecture in \cite{KDS}.
\begin{con}\cite{KDS}
The minimal graph for $I_{f^2}(G)$ with the exponential sequence is
a tree. Further it is a generalized star of diameter approximately
$\sqrt{2n}$ and, hence, with approximately $\sqrt{2n}$ branches.
\end{con}

Interestingly, the graph $9(linear)$ is also one of the maximal
graphs for $I_{f^\sigma}(G)$ in $N_9$, where $N_i$ is the set of all
non-isomorphic graphs on $i$ vertices. In addition, one elementary
result on maximal graphs with respect to $f^\sigma$ is also
obtained.
\begin{lem}\cite{KDS}
(i) A graph $G$ is maximal with respect to $I_{f^\sigma}(G)$ if and
only if its every vertex is an endpoint of a maximal path in $G$.

(ii) A maximal graph different from the complete graph $K_n$ cannot contain a vertex of degree $n-1$.
\end{lem}

\begin{figure}[h,t,b,p]
\begin{center}
\includegraphics[scale = 0.9]{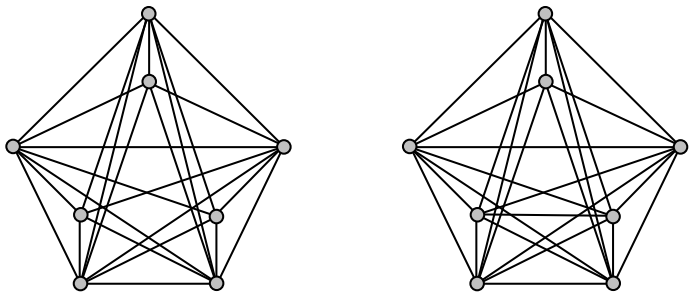}
\caption{Minimal graphs for $I_{f^\sigma}(G)$ of order $8$.}
\label{fig4}
\end{center}
\end{figure}

Similar to the case of $I_{f^2}(G)$, there is still very little
known about minimal entropy graphs respect to $I_{f^\sigma}(G)$. For
$N_8$ and $N_9$, computations show that there are $2$ minimal
graphs. For $n=8$, they are depicted in Figure \ref{fig4}, for $n=9$
they contain $5$ vertices of degree $8$ each. The authors \cite{KDS}
gave another conjecture as follows.

\begin{con}\cite{KDS}
A minimal graph for $I_{f^\sigma}(G)$ is a highly connected graph,
i.e., it is a graph obtained from the complete graph $K_n$ by
removal of a small number of edges. In particular, we conjecture
that a minimal graph for $I_{f^\sigma}(G)$ on $n$ vertices will have
$m\geq \frac{n}{2}$ vertices of degree $n-1$.
\end{con}

\subsection{Information inequalities for generalized graph entropies}

Sivakumar and Dehmer \cite{SD} discussed the problem of establishing
relations between information measures for network structures. Two
types of entropy measures, namely, the Shannon entropy and its
generalization, the R\'{e}nyi entropy have been considered for their
study. They established formal relationships, by means of
inequalities, between these two kinds of measures. In addition, they
proved inequalities connecting the classical partition-based graph
entropies and partition-independent entropy measures, and also gave
several explicit inequalities for special classes of graphs.

To begin with, we give the theorem which provide the bounds for
R\'{e}nyi entropy in terms of Shannon entropy.
\begin{thm}\label{thmBB}\cite{SD}
Let $p^f(v_1),p^f(v_2),\cdots,p^f(v_n)$ be the probability values on
the vertices of a graph $G$ with $n$ vertices. Then the R\'{e}nyi
entropy can be bounded by the Shannon entropy as follows:
\\[2mm]
when $0<\alpha<1$,
\begin{equation*}
I_f(G)\leq I^2_\alpha(G)_f<I_f(G)
+\frac{n(n-1)(1-\alpha)\rho^{\alpha-2}}{2\ln2},
\end{equation*}
when $\alpha>1$,
\begin{equation*}
I_f(G)-\frac{(\alpha-1)n(n-1)}{2\ln2\cdot\rho^{\alpha-2}}
<I^2_\alpha(G)_f\leq I_f(G),
\end{equation*}
where $\rho=max_{i,k}\frac{p^f(v_i)}{p^f(v_k)}$.
\end{thm}

Observe that Theorem \ref{thmBB}, in general, holds for any
probability distribution with non-zero probability values. The
following theorem illustrates this fact with the help of a
probability distribution obtained by partitioning a graph object.
\begin{thm}\cite{SD}
Let $p_1,p_2,\cdots,p_k$ be the probabilities of the partitions
obtained using an equivalence relation $\tau$ as stated before. Then
\\[2mm]
when $0<\alpha<1$,
\begin{equation*}
I(G,\tau)\leq I^1_\alpha(G)<I(G,\tau)
+\frac{k(k-1)(1-\alpha)\rho^{\alpha-2}}{2\ln2},
\end{equation*}
when $\alpha>1$,
\begin{equation*}
I(G,\tau)\geq I^1_\alpha(G)>I(G,\tau)
-\frac{k(k-1)(\alpha-1)}{2\ln2\cdot\rho^{\alpha-2}},
\end{equation*}
where $\rho=max_{i,j}\frac{p_i}{p_j}$.
\end{thm}

In the next theorem, bounds between like-entropy measures are
established, by considering the two different probability
distributions.
\begin{thm}\cite{SD}
Let $G$ be a graph with $n$ vertices. Suppose $|X_i|<f(v_i)$ for
$1\leq i\leq k$. Then
\begin{equation*}
I^1_\alpha(G)<I^2_\alpha(G)_f
+\frac{\alpha}{1-\alpha}\log_2\left(\frac{S}{|X|}\right)
\end{equation*}
if \ $0<\alpha<1$, and
\begin{equation*}
I^1_\alpha(G)>I^2_\alpha(G)_f
-\frac{\alpha}{\alpha-1}\log_2\left(\frac{S}{|X|}\right)
\end{equation*}
if \ $\alpha>1$. Here $S=\sum_{i=1}^{n}f(v_i)$.
\end{thm}

Furthermore, Sivakumar and Dehmer \cite{SD} also paid attention to
generalized graph entropies which is inspired by the R\'{e}nyi
entropy, and presented various bounds when two different functions
and their probability distributions satisfy certain initial
conditions. Let $f_1$ and $f_2$ be two information functions
defined on $G=(V,E)$ with $|V|=n$. Let $S_1=\sum_{i=1}^{n}f_1(v_i)$
and $S_2=\sum_{i=1}^{n}f_2(v_i)$. Let $p^{f_1}(v)$ and $p^{f_2}(v)$
denote the probabilities of $f_1$ and $f_2$, respectively, on a
vertex $v\in V$.
\begin{thm}\cite{SD}
Suppose $p^{f_1}(v)\leq \psi p^{f_2}(v)$, $\forall v\in V$ and
$\psi>0$ a constant. Then,
\\[2mm]
if \ $0<\alpha<1$,
\begin{equation*}
I^2_\alpha(G)_{f_1}\leq I^2_\alpha(G)_{f_2}+
\frac{\alpha}{1-\alpha}\log_2\psi,
\end{equation*}
and if \ $\alpha>1$,
\begin{equation*}
I^2_\alpha(G)_{f_1}\geq I^2_\alpha(G)_{f_2}-
\frac{\alpha}{\alpha-1}\log_2\psi.
\end{equation*}
\end{thm}

\begin{thm}\cite{SD}
Suppose $p^{f_1}(v)\leq p^{f_2}(v)+\phi$, $\forall v\in V$ and
$\phi>0$ a constant. Then,
\begin{equation*}
I^2_\alpha(G)_{f_1}-I^2_\alpha(G)_{f_2}<
\frac{1}{1-\alpha}\frac{n\cdot\phi^\alpha}{\sum\limits_{v\in
V}\left(p^{f_2}(v)\right)^\alpha}
\end{equation*}
if \ $0<\alpha<1$, and
\begin{equation*}
I^2_\alpha(G)_{f_2}-I^2_\alpha(G)_{f_1}<
\frac{\alpha}{\alpha-1}\frac{n^{\frac{1}{\alpha}}\cdot\phi}
{\left(\sum\limits_{v\in V}\left(p^{f_2}(v)\right)^\alpha\right)
^{\frac{1}{\alpha}}}
\end{equation*}
if \ $\alpha>1$.
\end{thm}

\begin{thm}\cite{SD}
Let $f(v)=c_1f_1(v)+c_2f_2(v)$, $\forall v\in V$. Then
\\[2mm]
for $0<\alpha<1$,
\begin{equation*}
I^2_\alpha(G)_{f}<I^2_\alpha(G)_{f_1}+
\frac{\alpha}{1-\alpha}\log_2A_1+\frac{1}{1-\alpha}
\frac{A^\alpha_2}{A^\alpha_1}
\frac{\sum\limits_{v\in V}\left(p^{f_2}(v)\right)^\alpha}
{\sum\limits_{v\in V}\left(p^{f_1}(v)\right)^\alpha},
\end{equation*}
and for $\alpha>1$,
\begin{equation*}
I^2_\alpha(G)_{f}>I^2_\alpha(G)_{f_1}-
\frac{\alpha}{\alpha-1}\log_2A_1-\frac{\alpha}{\alpha-1}
\frac{A_2}{A_1}\left(\frac{\sum\limits_{v\in V}
\left(p^{f_2}(v)\right)^\alpha}{\sum\limits_{v\in V}
\left(p^{f_1}(v)\right)^\alpha}\right)^{\frac{1}{\alpha}},
\end{equation*}
where $A_1=\frac{c_1S_1}{c_1S_1+c_2S_2}$ and $A_2=\frac{c_2S_2}{c_1S_1+c_2S_2}$.
\end{thm}

\begin{thm}\cite{SD}
Let $f(v)=c_1f_1(v)+c_2f_2(v)$, $\forall v\in V$. Then
\\[2mm]
if \ $0<\alpha<1$,
\begin{eqnarray*}
I^2_\alpha(G)_{f}&<&\frac{1}{2}\left[I^2_\alpha(G)_{f_1}
+I^2_\alpha(G)_{f_2}\right]+\frac{\alpha}{2(1-\alpha)}\log_2(A_1A_2)\\
& &+\frac{1}{2(1-\alpha)}\left[
\frac{A^\alpha_2}{A^\alpha_1}
\frac{\sum\limits_{v\in V}\left(p^{f_2}(v)\right)^\alpha}
{\sum\limits_{v\in V}\left(p^{f_1}(v)\right)^\alpha}
+\frac{A^\alpha_1}{A^\alpha_2}
\frac{\sum\limits_{v\in V}\left(p^{f_1}(v)\right)^\alpha}
{\sum\limits_{v\in V}\left(p^{f_2}(v)\right)^\alpha}\right],
\end{eqnarray*}
and if \ $\alpha>1$,
\begin{eqnarray*}
I^2_\alpha(G)_{f}&>&\frac{1}{2}\left[I^2_\alpha(G)_{f_1}
+I^2_\alpha(G)_{f_2}\right]-\frac{\alpha}{2(\alpha-1)}\log_2(A_1A_2)\\
& &-\frac{\alpha}{2(\alpha-1)}\left[
\frac{A_2}{A_1}\left(\frac{\sum\limits_{v\in V}
\left(p^{f_2}(v)\right)^\alpha}{\sum\limits_{v\in V}\left(p^{f_1}(v)\right)^\alpha}\right)^{\frac{1}{\alpha}}
+\frac{A_1}{A_2}\left(\frac{\sum\limits_{v\in V}
\left(p^{f_1}(v)\right)^\alpha}{\sum\limits_{v\in V}\left(p^{f_2}(v)\right)^\alpha}\right)^{\frac{1}{\alpha}}\right],
\end{eqnarray*}
where $A_1=\frac{c_1S_1}{c_1S_1+c_2S_2}$ and $A_2=\frac{c_2S_2}{c_1S_1+c_2S_2}$.
\end{thm}

Let $S_n$ be a star on $n$ vertices whose central vertex is denoted
by $u$. Let $\tau$ be an automorphism defined on $S_n$ such that
$\tau$ partitions $V(S_n)$ into two orbits, $V_1$ and $V_2$, where
$V_1=\{u\}$ and $V_2=V(S_n)-\{u\}$.
\begin{thm}\cite{SD}
If $\tau$ is the automorphism, as defined above, on $S_n$, then
\\[2mm]
for $0<\alpha<1$,
\begin{equation*}
I^1_\alpha(S_n)<\log_2n-\frac{n-1}{n}\log_2(n-1)
+\frac{(1-\alpha)(n-1)^{\alpha-2}}{\ln2},
\end{equation*}
and for $\alpha>1$,
\begin{equation*}
I^1_\alpha(S_n)>\log_2n-\frac{n-1}{n}\log_2(n-1)
-\frac{\alpha-1}{(n-1)^{\alpha-2}\ln2}.
\end{equation*}
\end{thm}

\begin{thm}\cite{SD}
Let $\tau$ be an automorphism on $V(S_n)$ and let $f$ be any
information function defined on $V(S_n)$ such that $|V_1|<f(v_i)$
and $|V_2|<f(v_j)$ for some $i,j$, $1\leq i\neq j\leq n$. Then
\\[2mm]
for $0<\alpha<1$,
\begin{equation*}
I^2_\alpha(S_n)_f>\frac{1}{1-\alpha}\log_2(1+(n-1)^\alpha)
-\frac{\alpha}{1-\alpha}\log_2S,
\end{equation*}
and for $\alpha>1$,
\begin{equation*}
I^2_\alpha(S_n)_f<\frac{1}{1-\alpha}\log_2(1+(n-1)^\alpha)
+\frac{\alpha}{\alpha-1}\log_2S,
\end{equation*}
where $S=\sum_{v\in V}f(v)$.
\end{thm}

The path graph, denoted by $P_n$, are the only trees with maximum
diameter among all the trees on $n$ vertices. Let $\tau$ be an
automorphism defined on $P_n$, where $\tau$ partitions the vertices
of $P_n$ into $\frac{n}{2}$ orbits $(V_i)$ of size $2$, when $n$ is
even, and $\frac{n-1}{2}$ orbits of size $2$ and one orbit of size
$1$, when $n$ is odd. Sivakumar and Dehmer \cite{SD} derived
equalities and inequalities on generalized graph entropies
$I^1_\alpha(P_n)$ and $I^2_\alpha(P_n)_f$ depending on the parity of
$n$.
\begin{thm}\cite{SD}
Let $n$ be an even integer and $f$ be any information function
such that $f(v)>2$ for at least $\frac{n}{2}$ vertices of $P_n$ and
let $\tau$ be stated as above. Then
\begin{equation*}
I^1_\alpha(P_n)=\log_2\frac{n}{2},
\end{equation*}
and
\begin{equation*}
I^2_\alpha(P_n)_f>\frac{1}{1-\alpha}\log_2n
-\frac{\alpha}{1-\alpha}\log_2S-1
\end{equation*}
if \ $0<\alpha<1$,
\begin{equation*}
I^2_\alpha(P_n)_f<\frac{1}{1-\alpha}\log_2n
+\frac{\alpha}{\alpha-1}\log_2S-1
\end{equation*}
if \ $\alpha>1$, where $S=\sum_{v\in V}f(v)$.
\end{thm}

\begin{thm}\cite{SD}
Let $n$ be an odd integer and $\tau$ be defined as before. Then
\\[2mm]
when $0\leq \alpha<1$,
\begin{equation*}
\log_2n-\frac{n-1}{n}\leq I^1_\alpha(P_n)<\log_2n+(n-1)\left[
\frac{(n+1)\cdot(1-\alpha)}{\ln2\cdot2^{5-\alpha}}-\frac{1}{n}\right],
\end{equation*}
and when $\alpha>1$,
\begin{equation*}
\log_2n-\frac{n-1}{n}\geq I^1_\alpha(P_n)>\log_2n-(n-1)\left[
\frac{(n+1)\cdot(\alpha-1)}{\ln2\cdot2^{\alpha+1}}+\frac{1}{n}\right].
\end{equation*}
Further if $f$ is an information function such that $f(v)>2$ for
at least $\frac{n+1}{2}$ vertices of $P_n$, then
\begin{equation*}
I^2_\alpha(P_n)_f>\frac{1}{1-\alpha}\log_2n
-\frac{\alpha}{1-\alpha}\log_2S-\frac{n-1}{n}
\end{equation*}
if \ $0<\alpha<1$, and
\begin{equation*}
I^2_\alpha(P_n)_f<\frac{1}{1-\alpha}\log_2n
+\frac{\alpha}{\alpha-1}\log_2S-\frac{n-1}{n}
\end{equation*}
if \ $\alpha>1$, where $S=\sum_{v\in V}f(v)$.
\end{thm}

In \cite{SD}, Sivakumar and Dehmer derived bounds of generalized
graph entropy $I^2_\alpha(G)_f$ for not only special graph classes
but also special information functions. Let $G=(V,E)$ be a simple
undirected graph on $n$ vertices and let $d(u,v)$, $S_j(u,G)$ be the
distance between $u,v$ and the $j$-sphere of $u$, respectively. For
the two special information functions $f_{P}(v_i)$ and
$f_{P'}(v_i)$, Sivakumar and Dehmer \cite{SD} presented the explicit
bounds for the graph entropy measures $I^2_\alpha(G)_{f_{P}}$ and
$I^2_\alpha(G)_{f_{P'}}$.
\begin{thm}\cite{SD}
Let $f_{P}$ be given by equation \ref{eqGG}. Let
$c_{max}=max\{c_i:1\leq i\leq \rho(G)\}$ and $c_{min}=min\{c_i:1\leq
i\leq \rho(G)\}$ where $c_i$ is defined in $f_{P}$. Then the value
of $I^2_\alpha(G)_{f_{P}}$ can be bounded as follows.
\\[2mm]
If \ $0<\alpha<1$,
\begin{equation*}
\log_2n-\frac{\alpha(n-1)X}{1-\alpha}\log_2\beta
\leq I^2_\alpha(G)_{f_{P}}\leq \log_2n
+\frac{\alpha(n-1)X}{1-\alpha}\log_2\beta,
\end{equation*}
and if \ $\alpha>1$,
\begin{equation*}
\log_2n-\frac{\alpha(n-1)X}{\alpha-1}\log_2\beta
\leq I^2_\alpha(G)_{f_{P}}\leq \log_2n
+\frac{\alpha(n-1)X}{\alpha-1}\log_2\beta,
\end{equation*}
where $X=c_{max}-c_{min}$.
\end{thm}

\begin{thm}\cite{SD}
Let $f_{P'}$ be given by equation \ref{eqHH}. Let
$c_{max}=max\{c_i:1\leq i\leq \rho(G)\}$ and $c_{min}=min\{c_i:1\leq
i\leq \rho(G)\}$ where $c_i$ is defined in $f_{P'}$. Then the value
of $I^2_\alpha(G)_{f_{P'}}$ can be bounded as follows.
\\[2mm]
If \ $0<\alpha<1$,
\begin{equation*}
\log_2n-\frac{\alpha}{1-\alpha}\log_2Y
\leq I^2_\alpha(G)_{f_{P'}}\leq\log_2n
+\frac{\alpha}{1-\alpha}\log_2Y,
\end{equation*}
and if \ $\alpha>1$,
\begin{equation*}
\log_2n-\frac{\alpha}{\alpha-1}\log_2Y
\leq I^2_\alpha(G)_{f_{P'}}\leq\log_2n
+\frac{\alpha}{\alpha-1}\log_2Y,
\end{equation*}
where $Y=\frac{c_{max}}{c_{min}}$.
\end{thm}

\section{Relationships between graph structures, graph energies, topological indices and generalized graph entropies}
\label{ch1:sec3}

In this section, we introduce ten generalized graph entropies based
on distinct graph matrices. Connections between such generalized
graph entropies and the graph energies, the spectral moments and
topological indices are provided. Moreover, we will give some
extremal properties of these generalized graph entropies and several
inequalities between them.

Let $G$ be a graph of order $n$ and $M$ be a matrix related to the
graph $G$. Denote $\mu_1, \mu_2, \cdots, \mu_n$ be the eigenvalues
of $M$ (or the singular values for some matrices). If $f:=|\mu_i|$,
then as defined in Definition \ref{defHH},
\begin{equation*}
p^f(v_i)=\frac{|\mu_i|}{\sum\limits_{j=1}^n|\mu_j|}.
\end{equation*}
Therefore, the generalized graph entropies are defined as follows:
\begin{eqnarray*}
&(i)& \
I^6(G)_\mu=\sum_{i=1}^n\frac{|\mu_i|}{\sum\limits_{j=1}^n|\mu_j|}
\left[1-\frac{|\mu_i|}{\sum\limits_{j=1}^n|\mu_j|}\right],\\
&(ii)& \ I^2_\alpha(G)_\mu=\frac{1}{1-\alpha}
\log\left(\sum_{i=1}^n\left(\frac{|\mu_i|}
{\sum\limits_{j=1}^n|\mu_j|}\right)^\alpha\right),\quad \alpha \neq
1,\\
&(iii)& \ I^4_\alpha(G)_\mu=\frac{1}{2^{1-\alpha}-1}\left(
\sum\limits_{i=1}^n\left(\frac{|\mu_i|}
{\sum\limits_{j=1}^n|\mu_j|}\right)^\alpha-1\right),\quad \alpha
\neq 1.
\end{eqnarray*}

1. Let $A(G)$ be the adjacency matrix of graph $G$ and the
eigenvalues of $A(G)$, $\lambda_1, \lambda_2, \cdots, \lambda_n$,
are said to be the eigenvalues of the graph $G$. The energy of $G$
is $\varepsilon(G)=\sum_{i=1}^{n}|\lambda_i|$. The $k$-th spectral
moment of the graph $G$ is defined as
$M_k(G)=\sum_{i=1}^{n}\lambda_i^k$. In \cite{ZJPRM}, the authors
defined the moment-like quantities,
$M^*_k(G)=\sum_{i=1}^{n}|\lambda_i|^k$.
\begin{thm}\cite{DLS}
Let $G$ be a graph with $n$ vertices and $m$ edges. Then for $\alpha\neq1$, we have
\begin{eqnarray*}
&(i)&\ I^6(G)_\lambda=1-\frac{2m}{\varepsilon^2},\\
&(ii)&\ I_\alpha^2(G)_\lambda=\frac{1}{1-\alpha}\log\frac{M_\alpha^*}
{\varepsilon^\alpha},\\
&(iii)&\ I_\alpha^4(G)_\lambda=\frac{1}{2^{1-\alpha}-1}
\left(\frac{M_\alpha^*}{\varepsilon^\alpha}-1\right),
\end{eqnarray*}
where $\varepsilon$ denotes the energy of graph $G$ and $M_\alpha^*=\sum\limits_{i=1}^n|\lambda_i|^\alpha$.
\end{thm}

The above theorem directly implies that for a graph $G$, each upper
(lower) bound of energy can be used to deduce an upper (a lower)
bound of $I^6(G)_\lambda$.
\begin{cor}\cite{DLS}

(i) \  For a graph $G$ with $m$ edges, we have
\begin{equation*}
\frac{1}{2}\leq I^6(G)_\lambda\leq 1-\frac{1}{2m}.
\end{equation*}

(ii) \  Let $G$ be a graph with $n$ vertices and $m$ edges. Then
\begin{equation*}
I^6(G)_\lambda\leq 1-\frac{1}{n}.
\end{equation*}

(iii) \ Let $T$ be a tree of order $n$. We have
\begin{equation*}
I^6(S_n)_\lambda\leq I^6(T)_\lambda\leq I^6(P_n)_\lambda,
\end{equation*}
where $S_n$ and $P_n$ denote the star graph and path graph of order $n$, respectively.

(iv) \ Let $G$ be a unicyclic graph of order $n$. Then we have
\begin{equation*}
I^6(G)_\lambda\leq I^6(P_n^6)_\lambda,
\end{equation*}
where $P_n^6$ \cite{LMW,LSWL} denotes the unicyclic graph obtained
by connecting a vertex of $C_6$ with a leaf of order $P_{n-6}$,
respectively.

(v) \ Let $G$ be a graph with $n$ vertices and $m$ edges. If its
cyclomatic number is $k=m-n+1$, then we have
\begin{equation*}
I^6(G)_\lambda\leq 1-\frac{2m}{(4n/\pi+c_k)^2},
\end{equation*}
where $c_k$ is a constant which only depends on $k$.
\end{cor}

2. Let $Q(G)$ be the signless Laplacian matrix of a graph $G$. Then
$Q(G)=D(G)+A(G)$, where $D(G)={\rm diag}(d_1, d_2, \cdots, d_n)$
denotes the diagonal matrix of vertex degrees of $G$ and $A(G)$ is
the adjacency matrix of $G$. Let $q_1, q_2, \cdots, q_n$ be the
eigenvalues of $Q(G)$.
\begin{thm}\cite{LQWGD}
Let $G$ be a graph with $n$ vertices and $m$ edges. Then for $\alpha \neq 1$, we have
\begin{eqnarray*}
&(i)& \ I^6(G)_q=1-\frac{1}{4m^2}(M_1+2m),\\
&(ii)& \ I_\alpha^2(G)_q=\frac{1}{1-\alpha}\log
\frac{M_\alpha^*}{(2m)^\alpha},\\
&(iii)& \ I_\alpha^4(G)_q=\frac{1}{2^{1-\alpha}-1}
\left(\frac{M_\alpha^*}{(2m)^\alpha}-1\right),
\end{eqnarray*}
where $M_1$ denotes the first Zagreb index and $M_\alpha^*=\sum\limits_{i=1}^n|q_i|^\alpha$.
\end{thm}

\begin{cor}\cite{LQWGD}

(i) \ For a graph $G$ with $n$ vertices and $m$ edges, we have
\begin{equation*}
I^6(G)_q \leq 1-\frac{1}{2m}-\frac{1}{n}.
\end{equation*}

(ii) \ Let $G$ be a graph with $n$ vertices and $m$ edges. The
minimum degree of $G$ is $\delta$ and the maximum degree of $G$ is
$\Delta$. Then
\begin{equation*}
I^6(G)_q\geq 1-\frac{1}{2m}-\frac{1}{2n}-\frac{\Delta^2+\delta^2}{4n\Delta\delta},
\end{equation*}
with equality if and only if $G$ is a regular graph, or $G$ is a
graph whose vertices have exactly two degrees $\Delta$ and $\delta$
such that $\Delta+\delta$ divides $\delta n$ and there are exactly
$p = \frac{\delta n}{\delta+\Delta}$ vertices of degree $\Delta$ and
$q=\frac{\Delta n}{\delta+\Delta}$ vertices of degree $\delta$.
\end{cor}

3. Let $\mathscr{L}(G)$ and $\mathcal {Q}(G)$ be the normalized
Laplacian matrix and the normalized signless Laplacian matrix,
respectively. By definition,
$\mathscr{L}(G)=D(G)^{-\frac{1}{2}}L(G)D(G)^{-\frac{1}{2}}$ and
$\mathcal {Q}(G)=D(G)^{-\frac{1}{2}}Q(G)D(G)^{-\frac{1}{2}}$, where
$D(G)$ is the diagonal matrix of vertex degrees, and
$L(G)=D(G)-A(G)$, $Q(G)=D(G)+A(G)$ are, respectively, the Laplacian
and the signless Laplacian matrices of the graph $G$. Denote the
eigenvalues of $\mathscr{L}(G)$ and $\mathcal {Q}(G)$ by $\mu_1,
\mu_2, \cdots, \mu_n$ and $q_1, q_2, \cdots, q_n$, respectively.

\begin{thm}\cite{LQWGD}
Let $G$ be a graph with $n$ vertices and $m$ edges. Then for $\alpha \neq 1$, we have
\begin{eqnarray*}
&(i)& \ I^6(G)_\mu=I^6(G)_q=1-\frac{1}{n^2}(n+2R_{-1}(G)),\\
&(ii)& \ I_\alpha^2(G)_\mu=
\frac{1}{1-\alpha}\log\frac{M_\alpha^*}{n^\alpha},\ \
I_\alpha^2(G)_q=\frac{1}{1-\alpha}\log\frac{M_\alpha^{*'}}{n^\alpha},\\
&(iii)& \ I_\alpha^4(G)_\mu=\frac{1}{2^{1-\alpha}-1}
\left(\frac{M_\alpha^*}{n^\alpha}-1\right),\ \
I_\alpha^4(G)_q=\frac{1}{2^{1-\alpha}-1}
\left(\frac{M_\alpha^{*'}}{n^\alpha}-1\right),
\end{eqnarray*}
where $R_{-1}(G)$ denotes the general Randi\'{c} index $R_\beta(G)$
of $G$ with $\beta=-1$ and
$M_\alpha^*=\sum\limits_{i=1}^n|\mu_i|^\alpha$, $M_\alpha^{*'}=\sum\limits_{i=1}^n|q_i|^\alpha$.
\end{thm}

\begin{cor}\cite{LQWGD}

(i) \ For a graph $G$ with $n$ vertices and $m$ edges, if $n$ is
odd, then we have
\begin{equation*}
1-\frac{2}{n}+\frac{1}{n^2}\leq I^6(G)_\mu=I^6(G)_q \leq 1-\frac{1}{n-1},
\end{equation*}
if $n$ is even, then we have
\begin{equation*}
1-\frac{2}{n}\leq I^6(G)_\mu=I^6(G)_q \leq 1-\frac{1}{n-1}
\end{equation*}
with right equality if and only if $G$ is a complete graph, and with
left equality if and only if $G$ is the disjoint union of
$\frac{n}{2}$ paths of length 1 for $n$ is even, and is the disjoint
union of $\frac{n-3}{2}$ paths of length 1 and a path of length 2
for $n$ is odd.

(ii) \ Let $G$ be a graph with $n$ vertices and $m$ edges. The
minimum degree of $G$ is $\delta$ and the maximum degree of $G$ is
$\Delta$. Then
\begin{equation*}
1-\frac{1}{n}-\frac{1}{n\delta}\leq I^6(G)_\mu=I^6(G)_q\leq 1-\frac{1}{n}-\frac{1}{n\Delta}.
\end{equation*}
Equality occurs in both bounds if and only if $G$ is a regular graph.
\end{cor}

4. Let $I(G)$ be the incidence matrix of a graph $G$ with vertex set
$V(G)=\{v_1, v_2, \cdots, v_n\}$ and edge set $E(G)=\{e_1, e_2,
\cdots, e_m\}$, such that the $(i,j)$-entry of $I(G)$ is 1 if the
vertex $v_i$ is incident with the edge $e_j$, and is 0 otherwise. As
we know, $Q(G)=D(G)+A(G)=I(G)\cdot I^T(G)$. If the eigenvalues of
$Q(G)$ are $q_1, q_2, \cdots, q_n$, then $\sqrt{q_1}, \sqrt{q_2},
\cdots, \sqrt{q_n}$ are the singular values of $I(G)$. In addition,
the incidence energy of $G$ is defined as $IE(G)=\sum_{i=1}^n
\sqrt{q_i}$.
\begin{thm}\cite{LQWGD}
Let $G$ be a graph with $n$ vertices and $m$ edges. Then for $\alpha \neq 1$, we have
\begin{eqnarray*}
&(i)& \ I^6(G)_{\sqrt{q}}=1-\frac{2m}{(IE(G))^2},\\
&(ii)& \ I_\alpha^2(G)_{\sqrt{q}}=\frac{1}{1-\alpha}\log
\frac{M_\alpha^*}{(IE(G))^\alpha},\\
&(iii)& \ I_\alpha^4(G)_{\sqrt{q}}=\frac{1}{2^{1-\alpha}-1}
\left(\frac{M_\alpha^*}{(IE(G))^\alpha}-1\right),
\end{eqnarray*}
where $IE(G)$ denotes the incidence energy of $G$ and $M_\alpha^*=\sum\limits_{i=1}^n(\sqrt{q_i})^\alpha$.
\end{thm}

\begin{cor}\cite{LQWGD}

(i) \ For a graph $G$ with $n$ vertices and $m$ edges, we have
\begin{equation*}
0 \leq I^6(G)_{\sqrt{q}} \leq 1-\frac{1}{n}.
\end{equation*}
The left equality holds if and only if $m\leq 1$, whereas the right equality holds if and only if $m = 0$.

(ii) \ Let $T$ be a tree of order $n$. Then we have
\begin{equation*}
I^6(S_n)_{\sqrt{q}} \leq I^6(T)_{\sqrt{q}} \leq I^6(P_n)_{\sqrt{q}},
\end{equation*}
where $S_n$ and $P_n$ denote the star and path of order $n$, respectively.
\end{cor}

5. Let the graph $G$ be a connected graph whose vertices are $v_1,
v_2, \cdots, v_n $. The distance matrix of $G$ is defined as $D(G) =
[d_{ij}]$, where $d_{ij}$ is the distance between the vertices $v_i$
and $v_j$ in $G$. We denote the eigenvalues of $D(G)$ by $\mu_1,
\mu_2, \cdots, \mu_n$. The distance energy of the graph $G$ is
$DE(G)=\sum_{i=1}^n|\mu_i|$.

The $k$-th distance moment of $G$ is defined as
$W_k(G)=\frac{1}{2}\sum_{1 \leq i <j \leq n}(d_{ij})^k$.
Particularly, $W(G)=W_1(G)$ and $WW(G)=\frac{1}{2}(W_2(G)+W_1(G))$,
where $W(G)$ and $WW(G)$ respectively denote the Wiener index and
hyper-Wiener index of $G$. We get the equality
$W_2(G)=\frac{1}{2}\sum_{1 \leq i <j \leq n}(d_{ij})^2=2WW(G)-W(G)$
by simple calculations. The following theorem describes the equality
relationships of the generalized graph entropy $I^6(G)_\mu$,
$DE(G)$, $W(G)$, $WW(G)$ and so on.
\begin{thm}\cite{LQWGD}
Let $G$ be a graph with $n$ vertices and $m$ edges. Then for $\alpha \neq 1$, we have
\begin{eqnarray*}
&(i)& \ I^6(G)_\mu=1-\frac{4}{(DE(G))^2}(2WW(G)-W(G)),\\
&(ii)& \ I_\alpha^2(G)_\mu=\frac{1}{1-\alpha}\log
\frac{M_\alpha^*}{(DE(G))^\alpha},\\
&(iii)& \ I_\alpha^4(G)_\mu=\frac{1}{2^{1-\alpha}-1}
\left(\frac{M_\alpha^*}{(DE(G))^\alpha}-1\right),
\end{eqnarray*}
where $M_\alpha^*=\sum_{i=1}^n|\mu_i|^\alpha$ and $DE(G)$ denotes
the distance energy of $G$. Here, $W(G)$ and $WW(G)$ are the Wiener
index and hyper-Wiener index of $G$, respectively.
\end{thm}

\begin{cor}\cite{LQWGD}
For a graph with $n$ vertices and $m$ edges, we have
\begin{equation*}
0 \leq I^6(G)_\mu\leq 1-\frac{1}{n}.
\end{equation*}
\end{cor}

6. Let $G$ be a simple undirected graph, and $G^\sigma$ be an
oriented graph of $G$ with the orientation $\sigma$. The skew
adjacency matrix of $G^\sigma$ is the $n \times n$ matrix
$S(G^\sigma)=[s_{ij}]$, where $s_{ij}=1$ and $s_{ji}=-1$ if $\langle
v_i,v_j\rangle$ is an arc of $G^\sigma$, otherwise
$s_{ij}=s_{ji}=0$. Let $\lambda_1, \lambda_2, \cdots, \lambda_n$ be
the eigenvalues of it. The skew energy of $G^\sigma$ is
$SE(G^\sigma)=\sum_{i=1}^n|\lambda_i|$.
\begin{thm}\cite{LQWGD}
Let $G^\sigma$ be an oriented graph with $n$ vertices and $m$ arcs. Then for $\alpha \neq 1$, we have
\begin{eqnarray*}
&(i)& \ I^6(G^\sigma)_\lambda=1-\frac{2m}{(SE(G^\sigma))^2},\\
&(ii)& \ I_\alpha^2(G^\sigma)_\lambda=\frac{1}{1-\alpha}
\log\frac{M_\alpha^*}{(SE(G^\sigma))^\alpha},\\
&(iii)& \ I_\alpha^4(G^\sigma)_\lambda=\frac{1}{2^{1-\alpha}-1}
\left(\frac{M_\alpha^*}{(SE(G^\sigma))^\alpha}-1\right),
\end{eqnarray*}
where $SE(G^\sigma)$ denotes the skew energy of $G^\sigma$ and $M_\alpha^*=\sum\limits_{i=1}^n|\lambda_i|^\alpha$.
\end{thm}

\begin{cor}\cite{LQWGD}

(i) \ For an oriented graph $G^\sigma$ with $n$ vertices, $m$ arcs
and maximum degree $\Delta$, we have
\begin{equation*}
1-\frac{2m}{2m+n(n-1)|det(S(G^\sigma))|^{\frac{2}{n}}} \leq I^6(G^\sigma)_\lambda\leq 1-\frac{1}{n} \leq 1-\frac{2m}{n^2\Delta}.
\end{equation*}

(ii) \ Let $T^\sigma$ be an oriented tree of order $n$. We have
\begin{equation*}
I^6(S_n^\sigma)_\lambda \leq I^6(T^\sigma)_\lambda \leq I^6(P_n^\sigma)_\lambda,
\end{equation*}
where $S_n^\sigma$ and $P_n^\sigma$ denote an oriented star and an
oriented path of order $n$ with any orientation, respectively.
Equality holds if and only if the underlying tree $T_n$ satisfies
that $T_n\cong S_n$ or $T_n \cong P_n$.
\end{cor}

7. Let $G$ be a simple graph. The Randi\'{c} adjacency matrix of $G$
is defined as $R(G)=[r_{ij}]$, where
$r_{ij}=(d_id_j)^{-\frac{1}{2}}$ if $v_i$ and $v_j$ are adjacent
vertices of $G$, otherwise $r_{ij}=0$. Denote $\rho_1, \rho_2,
\cdots, \rho_n$ be its eigenvalues. The Randi\'{c} energy of the
graph $G$ is defined as $RE(G)=\sum_{i=1}^n|\rho_i|$.
\begin{thm}\cite{LQWGD}
Let $G$ be a graph with $n$ vertices and $m$ edges. Then for $\alpha \neq 1$, we have
\begin{eqnarray*}
&(i)& \ I^6(G)_\rho=1-\frac{2}{(RE(G))^2}R_{-1}(G),\\
&(ii)& \ I_\alpha^2(G)_\rho=\frac{1}{1-\alpha}\log\frac{M_\alpha^*}
{(RE(G))^\alpha},\\
&(iii)& \ I_\alpha^4(G)_\rho=\frac{1}{2^{1-\alpha}-1}
\left(\frac{M_\alpha^*}{(RE(G))^\alpha}-1\right),
\end{eqnarray*}
where $RE(G)$ denotes the Randi\'{c} energy of $G$, and $R_{-1}(G)$
denotes the general Randi\'{c} index $R_{\beta}(G)$ of $G$ with
$\beta=-1$ and $M_\alpha^*=\sum_{i=1}^n|\rho_i|^\alpha$.
\end{thm}

\begin{cor}\cite{LQWGD}
For a graph with $n$ vertices and $m$ edges, we have
\begin{equation*}
I^6(G)_\rho \leq 1-\frac{1}{n}.
\end{equation*}
Equality is attained if and only if $G$ is the graph without edges, or if all its
vertices have degree one.
\end{cor}

8. Let $G$ be a simple graph with vertex set $V(G)=\{v_1, v_2,
\cdots, v_n\}$ and edge set $E(G)= \{e_1, e_2, \cdots, e_m\}$, and
let $d_i$ be the degree of vertex $v_i, i = 1, 2, \cdots, n$. Define
an $n \times m$ matrix whose $(i, j)$-entry is
$(d_i)^{-\frac{1}{2}}$ if $v_i$ is incident to $e_j$ and 0
otherwise. We call it the Randi\'{c} incidence matrix of $G$ and
denote it by $I_R(G)$. Obviously, $I_R(G)=D(G)^{-\frac{1}{2}}I(G)$.
Let $\sigma_1, \sigma_2, \cdots, \sigma_n$ be its singular values.
And also $\sum_{i=1}^n\sigma_i$ are defined as the Randi\'{c}
incidence energy $I_RE(G)$ of the graph $G$. Let $U$ be the set of
isolated vertices of $G$ and $W=V(G)-U$. Set $r=|W|$. Then we have
$\sum_{i=1}^n\sigma_i^2=r$. Particularly, $\sum_{i=1}^n
\sigma_i^2=n$ if $G$ has no isolated vertices.
\begin{thm}\cite{LQWGD}
Let $G$ be a graph with $n$ vertices and $m$ edges. Let $U$ be the
set of isolated vertices of $G$ and $W=V(G)-U$. Set $r=|W|$. Then
for $\alpha \neq 1$, we have
\begin{eqnarray*}
&(i)& \ I^6(G)_\sigma=1-\frac{r}{(I_RE(G))^2},\\
&(ii)& \ I_\alpha^2(G)_\sigma=\frac{1}{1-\alpha}\log\frac{M_\alpha^*}
{(I_RE(G))^\alpha},\\
&(iii)& \ I_\alpha^4(G)_\sigma=\frac{1}{2^{1-\alpha}-1}
\left(\frac{M_\alpha^*}{(I_RE(G))^\alpha}-1\right),
\end{eqnarray*}
where $I_RE(G)$ denotes the Randi\'{c} incidence energy of $G$ and
$M_\alpha^*=\sum\limits_{i=1}^n|\sigma_i|^\alpha$.
\end{thm}

\begin{cor}\cite{LQWGD}

(i) \ For a graph $G$ with $n$ vertices and $m$ edges, we have
\begin{equation*}
I^6(G)_\sigma\geq 1-\frac{r}{n},
\end{equation*}
the equality holds if and only if $G \cong K_2$.

(ii) \ Let $G$ be a graph with $n$ vertices and $m$ edges. Then
\begin{equation*}
I^6(G)_\sigma\leq 1-\frac{r}{n^2-3n+4+2\sqrt{2(n-1)(n-2)}},
\end{equation*}
the equality holds if and only if $G \cong K_n$.

(iii) \ Let $T$ be a tree of order $n$. We have
\begin{equation*}
I^6(T)_\sigma \leq I^6(S_n)_\sigma,
\end{equation*}
where $S_n$ denotes the star graph of order $n$.
\end{cor}

9. Let $R_\beta(G)$ be the general Randi\'{c} matrix of a graph $G$.
Define $R_\beta(G)=[r_{ij}]$, where $r_{ij}=(d_id_j)^{-\beta}$ if
$v_i$ and $v_j$ are adjacent vertices of $G$, otherwise $r_{ij}=0$.
Set $\gamma_1, \gamma_2, \cdots, \gamma_n$ be the eigenvalues of
$R_\beta(G)$. By the definition of $R_\beta(G)$ we can get
$R_\beta(G)=D(G)^\beta A(G)D(G)^\beta$ and
$\sum_{i=1}^n\gamma_i^2=tr(R_\beta^2(G))=2\sum_{i\sim
j}(d_id_j)^{2\beta}$ directly. The general Randi\'{c} energy is
defined as $RE_\beta(G)=\sum_{i=1}^n|\gamma_i|$. Similarly, we
obtain the theorem as follows.

\begin{thm}\cite{LQWGD}
Let $G$ be a graph with $n$ vertices and $m$ edges. Then for $\alpha \neq 1$, we have
\begin{eqnarray*}
&(i)& \ I^6(G)_\gamma=1-\frac{2}{(RE_\beta(G))^2}R_{2\beta}(G),\\
&(ii)& \ I_\alpha^2(G)_\gamma=\frac{1}{1-\alpha}
\log\frac{M_\alpha^*}{(RE_\beta(G))^\alpha},\\
&(iii)& \ I_\alpha^4(G)_\gamma=\frac{1}{2^{1-\alpha}-1}
\left(\frac{M_\alpha^*}{(RE_\beta(G))^\alpha}-1\right),
\end{eqnarray*}
where $RE_\beta(G)$ denotes the general Randi\'{c} energy of $G$,
and $R_{2\beta}(G)$ denotes the general Randi\'{c} index of $G$ and
$M_\alpha^*=\sum_{i=1}^n|\gamma_i|^\alpha$.
\end{thm}

10. Let $G$ be a simple undirected graph, and $G^\sigma$ be an
oriented graph of $G$ with the orientation $\sigma$. The skew
Randi\'{c} matrix of $G^\sigma$ is the $n \times n$ matrix
$R_s(G^\sigma)=[(r_s)_{ij}]$, where
$(r_s)_{ij}=(d_id_j)^{-\frac{1}{2}}$ and
$(r_s)_{ji}=-(d_id_j)^{-\frac{1}{2}}$ if $\langle v_i,v_j\rangle$ is
an arc of $G^\sigma$, otherwise $(r_s)_{ij}=(r_s)_{ji}=0$. Let
$\rho_1, \rho_2, \cdots, \rho_n$ be the eigenvalues of
$R_s(G^\sigma)$. It follows that
$R_s(G^\sigma)=D(G)^{-\frac{1}{2}}S(G^\sigma)D(G)^{-\frac{1}{2}}$
and $\sum_{i=1}^n\rho_i^2=tr(R_s^2(G^\sigma))=-2\sum_{i\sim
j}(d_id_j)^{-1}=-2R_{-1}(G)$, which implies that
$\sum_{i=1}^n|\rho_i|^2=2R_{-1}(G)$. The skew Randi\'{c}
energy is $RE_s(G^\sigma)=\sum_{i=1}^n|\rho_i|$.
\begin{thm}\cite{LQWGD}
Let $G^\sigma$ be an oriented graph with $n$ vertices and $m$ arcs.
Then for $\alpha \neq 1$, we have
\begin{eqnarray*}
&(i)& \ I^6(G^\sigma)_\rho=1-\frac{2}{(RE_S(G^\sigma))^2}R_{-1}(G),\\
&(ii)& \ I_\alpha^2(G^\sigma)_\rho=\frac{1}{1-\alpha}
\log\frac{M_\alpha^*}{(RE_S(G^\sigma))^\alpha},\\
&(iii)& \ I_\alpha^4(G^\sigma)_\rho=\frac{1}{2^{1-\alpha}-1}
\left(\frac{M_\alpha^*}{(RE_S(G^\sigma))^\alpha}-1\right),
\end{eqnarray*}
where $RE_S(G^\sigma)$ denotes the skew Randi\'{c} energy of
$G^\sigma$, and $R_{-1}(G)$ denotes the general Randi\'{c} index of
the underlying graph $G$ with $\beta=-1$ and
$M_\alpha^*=\sum_{i=1}^n|\rho_i|^\alpha$.
\end{thm}

\begin{cor}\cite{LQWGD}
For an oriented graph $G^\sigma$ with $n$ vertices and $m$ arcs, we have
\begin{equation*}
I^6(G^\sigma)_\rho \leq 1-\frac{1}{n}.
\end{equation*}
\end{cor}

For the above ten distinct entropies, we present the following
results on implicit information inequality, which can be obtained by
the method in \cite{DLS}.
\begin{thm}\cite{LQWGD}

(i) \ When $0<\alpha <1$, we have $I_\alpha^2 <I_\alpha^4\cdot
\ln2$; and when $\alpha>1$, we have $I_\alpha^2>
\frac{(1-2^{1-\alpha})\ln2}{\alpha-1}I_\alpha^4$.

(ii) \ When $\alpha \geq 2$ and $0<\alpha <1$, we have
$I_\alpha^4>I^6$; when $1<\alpha <2$, we have $I^6>
(1-2^{1-\alpha})I_\alpha^4$.

(iii) \ When $\alpha \geq 2$, we have $I_\alpha^2>
\frac{(1-2^{1-\alpha})\ln2}{\alpha-1}I^6;$ when $1<\alpha <2$, we
have $I_\alpha^2> \frac{(1-2^{1-\alpha})^2\ln2}{\alpha-1}I^6;$ when
$0<\alpha <1$, we have $I_\alpha^2>I^6$.
\end{thm}

\section{Summary and conclusion}
\label{ch1:sec4}

The entropy of a probability distribution can be interpreted not
only as a measure of uncertainty, but also as a measure of
information, and the entropy of a graph is an information-theoretic
quantity for measuring the complexity of a graph.
Information-theoretic network complexity measures have already been
intensely used in mathematical and medicinal chemistry including
drug design. So far, numerous such measures have been developed such
that it is meaningful to show relatedness between them.

This chapter mainly attempts to capture the extremal properties of
different (generalized) graph entropy measures and to describe
various connections and relationships between (generalized) graph
entropies and other variables in graph theory. The first section
aims to introduce various entropy measures contained in distinct
entropy measure classes. Inequalities and extremal properties of
graph entropies and generalized graph entropies, which are based on
different information functions or distinct graph classes, have
been described in Section 2. The last section focuses on the
generalized graph entropies and shows the relationships between
graph structures, graph energies, topological indices and some
selected generalized graph entropies. In addition, throughout this
chapter, we also state various applications of graph entropies
together with some open problems and conjectures for further
research.

Actually, graph entropy measures can be used to derive so-called
implicit information inequalities for graphs. Generally, information
inequalities describe relations between information measures for
graphs. In \cite{DM3}, the authors found and proved implicit
information inequalities which were also stated in the survey paper
\cite{DM1}. As a consequence, we will not give the detail results in
this aspect.

It is worth mentioning that many numerical results and analyses have
been obtained, which we refer the details to
\cite{DM2,Dehmer,ED,DBE,CDES,DM3, DMShi, DLLS}. These numerical
results imply that the change of different entropies corresponds to
different structural properties of graphs. Even for special graphs,
such as trees, stars, paths and regular graphs, the increase or
decrease of graph entropies implies special properties of these
graphs. As is known to all, graph entropy measures have important
applications in a variety of problem areas, including information
theory, biology, chemistry, and sociology, which we refer to
\cite{DBVG,DVBE,DBVG2,DSV,DE,HOS,ED2,ISG,SG} for details. This
further inspires researchers to explore the extremal properties and
relationships among these (generalized) graph entropies.

\end{document}